\documentclass[aps,prx,preprint,superscriptaddress,floatfix]{revtex4-2}

\usepackage[pdftex,colorlinks=true,hypertexnames=false]{hyperref}
\usepackage{amsbsy,amssymb,amsmath,bm}
\usepackage{graphicx,color,epsfig,rotate}
\usepackage{float}
\usepackage{xspace}
\usepackage{placeins}
\usepackage{ulem}
\usepackage{xcolor}

\newcommand {\UTe}{UTe$_\mathrm{2}$\xspace}
\newcommand {\Hm}{$H_\mathrm{m}$\xspace}
\newcommand {\Hc}{$H_\mathrm{c2}$\xspace}
\newcommand {\Tc}{$T_\mathrm{c}$\xspace}
\newcommand {\He}{$H_\mathrm{ex}$\xspace}

\begin{document}

\title{Field-induced compensation of magnetic exchange as the possible origin of reentrant superconductivity in \texorpdfstring{\UTe}{UTe2}}

\author{Toni Helm$^{1,2,\ast}$, Motoi Kimata$^{3}$, Kenta Sudo$^{3}$, Atsuhiko Miyata$^{1}$,\\
Julia Stirnat$^{1,4}$, Tobias F\"orster$^{1}$, Jacob Hornung$^{1,4}$,\\
Markus K\"onig$^{2}$, Ilya Sheikin$^{5}$, Alexandre Pourret$^{6}$,Gerard Lapertot$^{6}$,
Dai Aoki$^{7}$, Georg Knebel$^{6}$, Joachim Wosnitza$^{1,4}$, Jean-Pascal Brison$^{6}$\\
{$^{1}$Hochfeld-Magnetlabor Dresden (HLD-EMFL) and W\"urzburg-Dresden Cluster of Excellence ct.qmat, Helmholtz-Zentrum Dresden-Rossendorf, 01328 Dresden, Germany}\\
{$^{2}$Max Planck Institute for Chemical Physics of Solids, 01187 Dresden, Germany}\\
{$^{3}$Institute for Materials Research, Tohoku University, Sendai, Miyagi, 980-8577, Japan}\\
{$^{4}$Institut f\"ur Festk\"orper- und Materialphysik, Technische Universit\"at Dresden, 01062 Dresden, Germany}\\
{$^{5}$Laboratoire National des Champs Magn\'etiques Intenses (LNCMI-EMFL), CNRS, UGA, 38042 Grenoble, France}\\
{$^{6}$Univ. Grenoble Alpes, CEA, Grenoble-INP, IRIG, PHELIQS, 38000 Grenoble, France}\\
{$^{7}$Institute for Materials Research, Tohoku University, Oarai,
Ibaraki, 311-1313, Japan}\\
{$^\ast$Corresponding author: \textbf{t.helm@hzdr.de}}\\
}
\date{\today}

\begin{abstract}

The potential spin-triplet heavy-fermion superconductor \UTe exhibits signatures of multiple distinct superconducting phases. 
For field aligned along the $b$ axis, a metamagnetic transition occurs at $\mu_0$\Hm$\approx35\,$T. It is associated with magnetic fluctuations that may be beneficial for the field-reinforced superconductivity surviving up to \Hm. 
Once the field is tilted away from the $b$ towards the $c$ axis, a reentrant superconducting phase emerges just above \Hm. 
In order to better understand this remarkably field-resistant superconducting phase, we conducted magnetic-torque and magnetotransport measurements in pulsed magnetic fields. 
We determine the record-breaking upper critical field of $\mu_0$\Hc$\approx 73\,$T and its evolution with angle. 
Furthermore, the normal-state Hall effect experiences a drastic suppression indicative of a reduced band polarization above \Hm in the angular range around $30^\circ$ caused by a partial compensation between the applied field and an exchange field. This promotes the Jaccarino-Peter effect as a likely mechanism for the reentrant superconductivity above \Hm. 
\end{abstract}

\maketitle

\section*{Introduction}
Superconductivity is notoriously fragile under magnetic field, all the more when the superconducting critical temperature is small. 
However, the sensitivity of superconductors to magnetic field is influenced by a variety of factors. 
For example, a whole class of strongly correlated electron systems called "heavy fermions" exhibits critical fields several orders of magnitude larger than other superconducting systems with similar \Tc (usually sub-Kelvin), precisely because the quasi particles possess heavy effective masses, or equivalently, very slow Fermi velocities\ \cite{Brison1997,Pfleiderer2009,Onuki2018}.
In many heavy-fermion materials, the upper critical field is limited at low temperatures by the paramagnetic limit that arises from the Zeeman coupling of the Cooper pair spins to the external field\ \cite{Clogston1962,Chandrasekhar1962}. 
In other superconductors, only a strong 2D character may allow for enhanced upper critical fields close to that limit. 

The recent discovery of superconductivity (SC) in the heavy-fermion metal \UTe\ \cite{RanScience2019} with a critical temperature \Tc$\approx 2\,$K, triggered much excitement, as its critical field reaches values approaching those of high-Tc superconductors. 
Moreover, \UTe appeared very quickly as a potential candidate for topological spin-triplet SC\ \cite{RanScience2019,AokiJPSJ2019} with multiple unconventional superconducting phases under field or pressure\ \cite{Nakamine2019, Fujibayashi2022OrderParam, JiaoNature2020, Hayes2021, Ran2019extreme, BraithwaiteCommPhys2019, LinNPJ2020, AokiJPSJ2020, ThomasPRB2021, KnafoCommPhys2021, Aoki2021Pressure, RosuelPRX2023}. 
Spin-triplet SC is a rare phenomenon, expected to arise as a consequence of magnetic fluctuations in strongly correlated materials. 
It is characterized by a particularly high stability against external magnetic fields due to the suppression of Pauli depairing. 
Indeed, a key characteristic of \UTe is an anisotropic upper critical field, \Hc, that exceeds the paramagnetic limit along all field orientations\ \cite{RanScience2019,Knebel2019}. 
In particular, SC survives up to a metamagnetic transition at approximately $\mu_0$\Hm$=35\,$T, for field oriented along the magnetically hard $b$ direction\ \cite{Knebel2019, Ran2019extreme,  RosuelPRX2023}. 
These findings resemble those reported for ferromagnetic superconductors, such as UCoGe and URhGe\ \cite{Aoki2009UCoGe,Levy2005URhGe}. 
More surprising, the compound is able to reestablish SC even at higher fields, just above $\mu_0$\Hm at $\sim40\,$T for field oriented at $\theta\approx 30^\circ$ away from $b$ towards the $c$ axis\ \cite{Ran2019extreme}. 

The new reentrant high-field superconducting phase (from here on referred to as hfSC phase) appears to extend into an extreme field range beyond $60\,$T, with a yet-to-be-determined \Hc\ \cite{Ran2019extreme,KnafoCommPhys2021}. 
The nature of the superconducting ground state, the identification of the different field or pressure-induced superconducting phases, and their relation to topological SC are still under debate, notably from a theoretical point of view\ \cite{Xu2019FermiSurface,IshizukaPRL2019,Shishidou2021TologigicalBand,IshizukaPRB2021}. 
The mechanisms behind this record-breaking hfSC phase and its relation to the low-field superconductivity (lfSC) are a puzzle and one of the key questions to solve.

Indeed, little is known about the mechanisms responsible for the high-field superconducting phases: 
neither is it clear how exactly SC is suppressed for $H \parallel b$ once \Hm is approached; nor why SC can reestablish for field orientations near $\approx 30^{\circ}$ within the ($b,c$) plane above \Hm. 
Hall-resistivity measurements with $H\parallel b$ revealed a significant anomalous Hall effect (AHE), which has been associated with coherent skew scattering that dominates the electrical transport below $T\approx 20\,$K\ \cite{NiuPRL2020}. 
A sign change in the ordinary Hall coefficient and thermoelectric power, and a discontinuity in the $T^2$ term of the temperature-dependent resistivity or the specific heat at \Hm indicate a strong impact of the metamagnetism on the electronic band structure
 and on the correlations\ \cite{NiuPRL2020,KnafoCommPhys2021, MiyakeA2021, RosuelPRX2023}. 

The field-reinforced SC observed for $H\parallel b$ below \Hm\ \cite{RanScience2019,Knebel2019} is associated with an enhancement of magnetic fluctuations in the vicinity of the metamagnetic transition\ \cite{Ran2019extreme, Knebel2019, KnafoJPSJ2019, Imajo2019, Miyake2019, RosuelPRX2023}.
Although \Hm represents the limiting scale for SC for $H\parallel b$, it is also the enabling lower barrier for the hfSC phase. 
This suggests that magnetic interactions connected with \Hm play a key role for the emergence of the field-enhanced and reentrant SC of \UTe. 
At low temperature, the metamagnetic signature is a step-like change in the magnetization\ \cite{Ran2019extreme, Miyake2019, MiyakeA2021} and in various transport properties such as the residual resistivity\ \cite{KnafoJPSJ2019, LinNPJ2020, KnafoCommPhys2021}, and the Hall effect\ \cite{NiuPRL2020}.
The metamagnetic transition is sensitive to the field alignment\ \cite{Ran2019extreme, KnafoCommPhys2021, MiyakeA2021}. 
It shifts to higher fields upon changing the field orientation either from $b$ to $c$ or from $b$ to $a$. 
However, the jump of the magnetization at \Hm seems to remain unaffected by an orientation change of $30^{\circ}$ within the $(b,c)$ plane\ \cite{Miyake2019, Ran2019extreme, MiyakeA2021}.
Presently, the only quantity that differs is the sign of the specific-heat jump at \Hm, negative for $H \parallel b$\ \cite{MiyakeA2021,RosuelPRX2023}, but becoming positive at $30^{\circ}$\ \cite{MiyakeA2021}. 
Interestingly, pressure-dependent investigations have revealed that the hfSC phase is not necessarily tied to \Hm: 
At large enough pressures, hfSC emerges at field values larger than \Hm\ \cite{Ran2021expansion}. 

Here , we present studies of magnetic torque, magnetoresistance, and Hall effect in pulsed magnetic fields up to $70\,$T for micron-sized samples. 
They are cut from single crystals of \UTe by focused-ion-beam (FIB) microfabrication.
This enables us to perform measurements in pulsed magnetic fields with enhanced precision in a rather noisy environment compared to steady magnetic fields. 
We trace the metamagnetic and superconducting transitions in the $(b,c)$ plane. 
We confirm the emergence of hfSC phase around $\theta = 30^\circ$ at fields above $40\,$T. 
We extrapolate the maximum upper critical field to $\mu_0$\Hc$\approx 73\,$T and determine its variation with angle. 
We trace the magnetic torque trough \Hm and demonstrate that the spins reside in a non-collinear configuration with a dominant $b$-axis component.   
Furthermore, we show that the high-field Hall coefficient, having an orbital and a significant AHE component, experiences a drastic suppression as the field orientation approaches the hfSC region around $30^\circ$, even though magnetization, magnetic torque and magnetoresistance remain finite. 
We propose a new interpretation of the AHE at low temperature in the polarized phase of \UTe above \Hm, 
which suggests a  scenario connecting the suppression of the AHE around $30^\circ$ and the emergence of the hfSC phase. 
It relies on a field-induced enhancement of the pairing strength together with an angle-dependent band polarization.  

\section*{Results and Discussion}

\begin{figure}[tb]
	\centering
	\includegraphics[width=0.9\textwidth]{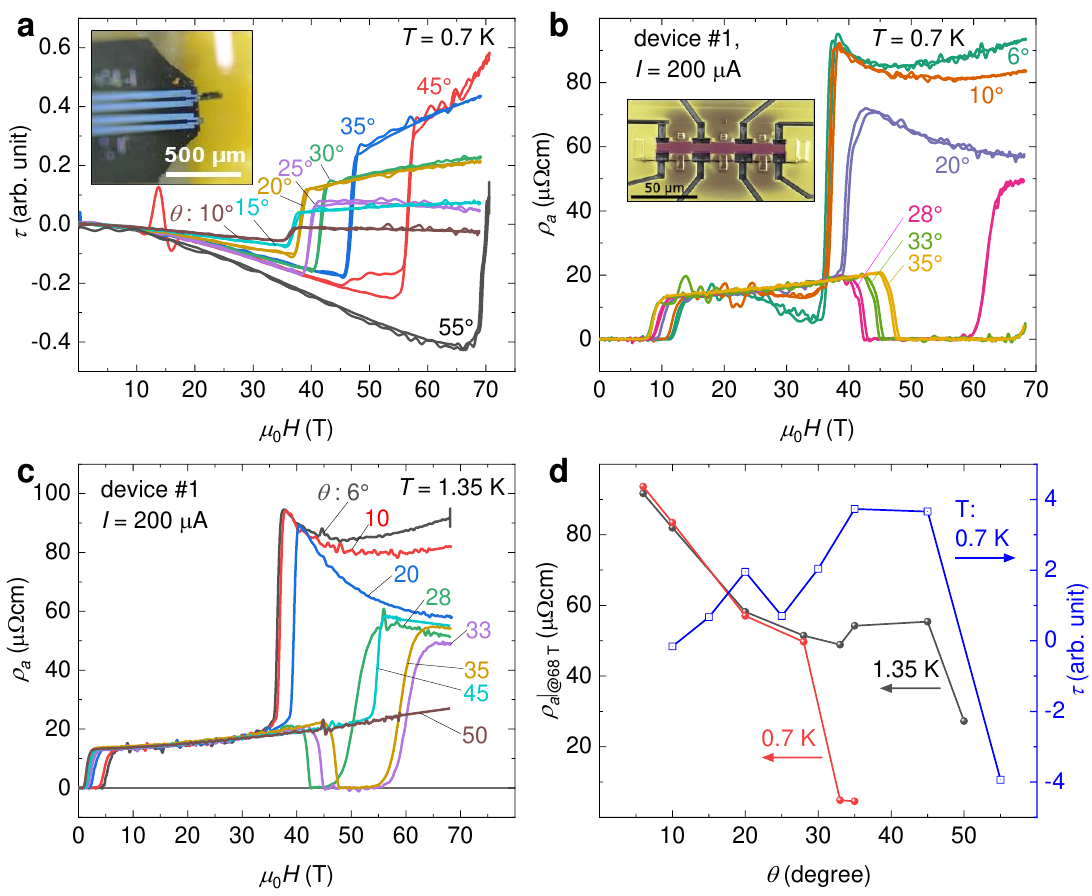}
	\caption{
    \textbf{Magnetic torque and magnetoresistivity of UTe$_2$.}
    (a) Magnetic torque vs. pulsed magnetic field for various angles recorded for a thin sample $(90\times 15\time 2)\,\mu$m at a temperature of $0.7\,$K. 
	The tilt angle, $\theta$, denotes the field orientation in the $(b,c)$ plane, where $0^\circ$ corresponds to $H\parallel b$. 
	Inset: Picture of the piezoresistive microcantilever with a lamella-shaped sample attached to it. 
	(b), (c) Resistivity vs. pulsed magnetic field for device $\#1$ recorded at $T=0.7$ and $1.35\,$K, respectively, for various tilt angles. 
    Inset of (b): False-color scanning-electron-microscope image of the FIB structured Hall-bar device $\#1$ with a thickness of $2\,\mu$m and $I \parallel a$. 
    The $b$ axis points along the normal of the substrate. 
    (d) First layer: Resistivity at $68\,$T from the data in (b) and (c) versus angle at $0.7$ and 1.35\,K. Second layer: Magnetic torque at $60\,$T from data in (a) versus angle at $0.7\,$K.
    }
	\label{fig1}
\end{figure}

We investigated several micron-sized samples produced from one oriented single crystal with a superconducting \Tc of 1.6\,K. 
The micromachining was performed by means of Ga or Xe FIB systems (for details, see the methods section).
This FIB approach enables precise geometries suitable for microcantilever-torque experiments on magnetic materials with strong torque responses as well as high-precision electrical-transport measurements on metallic (i.e., highly conductive) materials with current running along any desired direction (see images in Fig.\ \ref{fig1}a and b). 
In this work, we will present results obtained for three transport devices shaped in the standard Hall-bar geometry. 
Additional torque and magnetotransport data are provided in Supplementary Notes\ S1-S3. 
A preliminary characterization of the zero-field resistivity in micron-sized structures yielded no significant differences compared to results reported for bulk samples, see Supplementary Fig.\ S2. 
The critical temperature of 1.6\,K is not altered by the fabrication in comparison to the bulk sample, and the overall temperature dependence is reproduced. 

\subsection*{Magnetic torque around the metamagnetic transition}

We investigated the isothermal magnetic torque of \UTe by means of microcantilever torque magnetometry (see Fig.\ \ref{fig1}a) in pulsed fields up to 70\,T for various angles. 
This technique probes the magnetic anisotropy and complements magnetization measurements\ \cite{Ohmichi2002}. 
As a consequence of the step-like increase of magnetization and the change in anisotropy of \UTe at \Hm\ \cite{Miyake2019,MiyakeA2022}, the response in magnetic torque is strong. Thanks to the sample preparation by FIB, the volume of the sample is small enough to limit the maximum torque to a safe value preventing damage to the microcantilever. 

Figure\ \ref{fig1}a presents torque data recorded at $0.7\,$K. 
An additional data set for $T = 1.5\,$K can be found in the Supplementary Information Fig.\ S1. 
The tilt angle $\theta$ was varied between $H\parallel b$, i.e., $\theta=0^\circ$, and the $c$ direction. 
The metamagnetic high-field transition shows up as a step-like feature at fields above $35\,$T. 
The monotonic change in $\tau(H)$ at constant angle reflects that of the bulk magnetization. 
It confirms that besides the jump at \Hm, there are no other anomalies in the magnetization for all the measured angles within the $(b,c)$ plane. 
Interestingly, the jump in $\tau(\theta)$ depending on the tilt angle exhibits a pronounced local minimum at $\theta \approx 25^\circ$, for all fields above \Hm in this angular range. 
This is best seen when we plot the torque magnitude against the tilt angle, at low temperature, see Fig.\ \ref{fig1}d and Supplementary Fig.\ S1. 
As the magnetic torque reflects the magnetic moment of the anisotropic crystal, it is sensitive to the magnetization component perpendicular to the magnetic field. 
Its maximum is expected around $45^\circ$, consistent with our data. 
The noticeable drop at $25^\circ$, therefore, is indicative of a bulk feature in the magnetic part of \UTe around this angular range coinciding with the reentrant high-field superconducting phase. 
However, at higher temperature than $T=1.5\,$K the feature seems absent, as can be seen in Supplementary Fig.\ S1. 
Note: Both the low-field and the high-field SC is not discernible in the pulsed-field torque data. This may be a consequence of the fast $dH/dt$.
The observed, drop of the torque around $25^\circ$, however, may originate from a screening associated with superconducting diamagnetic currents. 
More work is required to pinpoint the origin of the decrease of the torque in this angular range.

Remarkably, the jump of the torque at \Hm for finite tilt angles changes from strictly negative to strictly positive values, not from negative values to zero. 
Therefore, in the ``polarized state'' above \Hm the magnetic moments and $H$ are not collinear and have a dominant $b$-axis component even for $\theta\geq45^\circ$. 
This is an additional feature revealed by our magnetic torque measurements. 
In comparison to the previous magnetization studies\ \cite{Ran2019extreme, Miyake2019}, magnetic torque is sensitive to the transverse component of the magnetization. 

\subsection*{High-field superconductivity and its electrical-transport signature}

We conducted resistivity and Hall-effect measurements in fields up to $70\,$T. 
Isothermal resistivity curves recorded for Hall-bar device $\#1$ (see inset in Fig.\ \ref{fig1}b) at $0.7$ and $1.3\,$K, with field oriented along the $b$ axis, are presented in Fig.\ \ref{fig1}b and c, respectively. 
The in-plane resistivity $\rho_{a}$ exhibits a step-like change at the metamagnetic transition that sets in at $\mu_0 H\approx 35\,$T for $H\parallel b$. 
This feature is consistent with the metamagnetic jump at \Hm in magnetic torque. 
We provide additional data recorded for device $\#3$ at various temperatures ranging between 1.4 and 77\,K for the fixed field orientation $H\parallel b$ in Supplementary Fig.\ S2. 
Upon decreasing temperature, the metamagnetic transition evolves from a broad anomaly into a sharp first-order type transition. 
Such an evolution resembles the behavior observed in other heavy-fermion metals with metamagnetism in high fields\ \cite{Settai_1997,Sakon_2002,Hamann2021UN}. 
Regarding the hfSC phase, our results are in line with previous reports\ \cite{Ran2019extreme, KnafoCommPhys2021}.
However, we show its extent to higher fields with a far improved resolution.

First, we focus on the data recorded for orientations close to $H\parallel b$: In the $6^\circ$ curve we observe a fingerprint of the reentrant behavior of the lfSC phase reported previously\ \cite{Knebel2019, RanScience2019}: 
the normal state is reached above $12\,$T until the resistivity starts dropping again above $20\,$T, see Fig.\ \ref{fig1}b. 
Apparently, the reentrant signature is suppressed in the $1.35\,$K data, shown in Fig.\ \ref{fig1}c. 
As we increase $\theta$ to $20^\circ$ and above, the magnetoresistance in the normal state below $H_\mathrm{M}$ remains unchanged. 
Above $H_\mathrm{M}$, it gradually evolves from a positive upturn into a monotonic change. 
Similar to our observations in magnetic torque, the metamagnetic transition shifts towards higher fields. 
However, in the case of resistivity, the strong step-like feature is quenched by the onset of zero resistance associated with an additional reentrant phase that sets in, once $\theta$ reaches beyond $20^\circ$. 
At $0.7\,$K, the resistivity curve for the highest tilt angles $\theta=35^\circ$ tested, still exhibits SC that extends up to $69\,$T. 
In comparison, for the same angle but at $1.3\,$K, the resistance reaches the normal state again already at fields below 60\,T. 
At $1.35\,$K, no trace of hfSC was discernible for angles from $45^\circ$ onward, see Fig.\ \ref{fig1}c. 
At $45^\circ$, we observe a step-like resistance increase followed by a negative slope as for angles below $28^\circ$. 
For $\theta=50^\circ$, $H_\mathrm{M}$ is pushed above the field range accessed in this experiment. 
Hence, the normal-state resistance increases monotonically up to the highest field.
As can be noticed from Fig.\ \ref{fig1}d, the magnetoresistsivity $\rho_a(\theta)$ in the normal state above \Hm experiences a slope change from positive to negative upon rotation away from $H\parallel b$. 
The overall amplitude at 68\,T exhibits a dip near $30\,^\circ$, the angle where the hfSC appears to be the strongest. 
 
\begin{figure}[tbh]
	\centering
	\includegraphics[width=.85\textwidth]{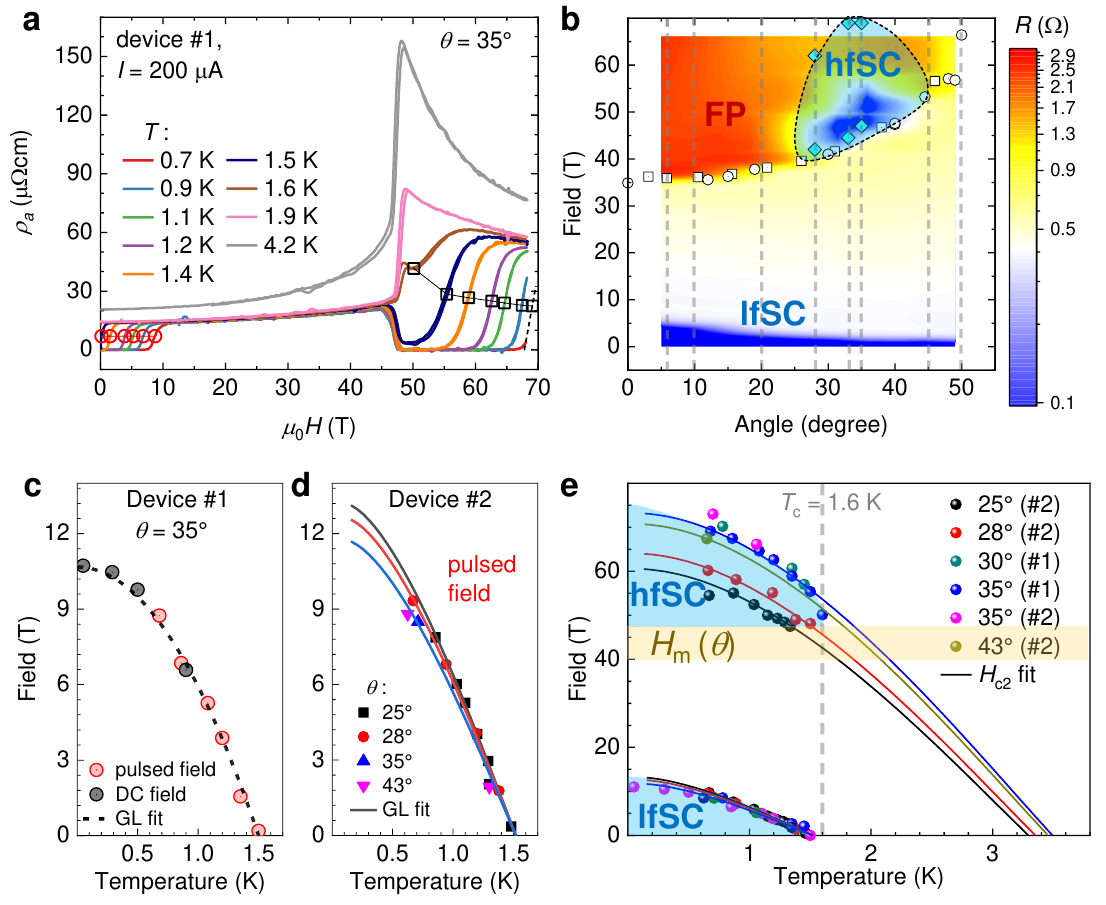}
	\caption{
    \textbf{High-field phase diagram of \UTe.}
    (a) $a$-axis resistivity versus magnetic field at fixed tilt angle of $\theta= 35^\circ$ for various temperatures. 
    (b) Contour plot created from data presented in Fig.~\ref{fig1}b and c. 
	White squares and circles mark the metamagnetic transition field measured by pulsed-field torque magnetometry presented in Fig.\ \ref{fig1}a and Fig.\ S1. 
    Cyan diamonds indicate the superconducting transition fields in the 0.7\,K data set presented in Fig.\ \ref{fig1}b. 
	The black dashed line indicates the approximate extension of the SC region at 0.7\,K.
    (c), (d) Temperature dependence of the superconducting critical field of the lfSC phase, determined for device $\#1$ and $\#2$, respectively, in pulsed and steady fields. Dashed and solid lines are GL fits (for details see Supplementary Note~5).  
	(e) Critical fields of the hfSC and lfSC phase determined in pulsed field. Solid lines are fits of \Hc in the pure orbital limit, using a strong-coupling constant $\lambda=1.51$, $1.53$, $1.58$, and  $1.567$, respectively, at $25^{\circ}$, $28^{\circ}$, $35^{\circ}$, and  $43^{\circ}$. 
    }
	\label{fig2}
\end{figure}

In Fig.\ \ref{fig2}a, we present a data set of the resistance recorded for various temperatures between 0.7 and $1.4\,$K at the fixed orientation $\theta=35^\circ$. 
The critical fields of the lfSC and the hfSC phases at different temperatures were determined from the inflection points of the magnetoresistance curves. 
At the lowest temperature reached in our experiment, $T = 0.7\,$K, the superconducting phase survives magnetic fields close to 70\,T. 
Its onset appears to be directly pinned to the metamagnetic transition.

\subsection*{\texorpdfstring{\Hc}{Hc2} in the field-induced reentrant hfSC phase}

Figure\ \ref{fig2}b shows a schematic phase diagram comprised of a contour presentation of the data from Fig.\ \ref{fig1}c and the transition fields $H_\mathrm{M}$ and $H_\mathrm{c}$ determined from our torque and resistance results.

Figure\ \ref{fig2}c shows the superconducting \Hc of the lfSC phase for device $\#1$ determined in DC (gray) and pulsed (red) magnetic fields oriented parallel to the $b$ axis (squares) and tilted $30^\circ$ (circles) towards the $c$ axis within the $(b,c)$ plane. 
Figure\ \ref{fig2}d shows similar data for device $\#2$ for fields applied at different angles within the $(b,c)$ plane, measured all in pulsed fields.

For spin-singlet superconductors, \Hc has an upper limit fixed by Pauli paramagnetism\ \cite{Clogston1962, Chandrasekhar1962}. 
The limiting field, $H_\mathrm{Pauli}$, for a singlet superconductor at 0\,K can be approximated by $\mu_0 H_\mathrm{Pauli}\approx \sqrt{2}\Delta/(g\mu_\mathrm{B}) = 1.86[\mathrm{T/K}]\cdot T_\mathrm{c}$, valid in the BCS weak-coupling limit without any spin-orbit interaction and for a free-electron value of the $g$-factor: $g=2$. 
In the case of \UTe, this would roughly lead to $3\,$T, much smaller than the measured critical fields (reaching close to or beyond 10\,T in all directions). 
A combination of spin-triplet SC, strong superconducting coupling, and strong spin-orbit interactions could be responsible for this violation of the paramagnetic limit in all field directions\ \cite{HiranumaJPSJ2021}.

The evolution of \Hc with temperature in the lfSC phase for the field tilted by approximately $35^\circ$ towards the $c$ axis follows the standard (close to parabolic) temperature dependence of an \Hc in the pure orbital limit. 
Fits of the data were done in the strong-coupling regime appropriate for \UTe\ \cite{AokiJPCM2022}, using a moderate value of the strong-coupling constant of $\lambda = 1$ (solid lines in Fig.\ \ref{fig2}c and d). 
In the Ginzburg-Landau weak-coupling regime, the slope of \Hc at \Tc is given by\ \cite{Bulaevskii1988Hc2}:
\begin{equation}\frac{dH_\mathrm{c2}}{dT} \approx 9 \Phi_0 \left(\frac{k_\mathrm{B}T_\mathrm{c}}{\hbar \langle v_\mathrm{F}\rangle}\right)^2 \frac{1}{T_\mathrm{c}}.
\end{equation}
Once $\lambda$ and \Tc are fixed, the slope of \Hc at \Tc (hence, the average Fermi velocity perpendicular to the applied field $\langle v_\mathrm{F}\rangle$) is the only parameter left to determine the complete temperature dependence of \Hc in this approximation. 
From the best fits, we find $6700\,\mathrm{m}/\mathrm{s}\leq\langle v_\mathrm{F}\rangle\leq7100\,\mathrm{m}/\mathrm{s}$
for angles between 25$^\circ$ and 35$^\circ$ in the $(b,c)$ plane. 

Let us now turn to the critical fields of the hfSC phase, above \Hm. 
The points shown in Fig.\ \ref{fig2}e were determined in the hfSC phase for two devices, again at various tilt angles. 
In our pulsed-field setup, we were limited to temperatures above $0.7\,$K. 
A prerequisite to a profound analysis of \Hc in this phase is a theoretical model explaining the mechanisms for reentrant SC above \Hm. 
Indeed, in the likely case of a connection between hfSC and magnetic fluctuations that develop upon approaching \Hm, we can expect a reduction of the pairing strength (following the observed reduction in the specific heat), $\lambda$, once the external magnetic field becomes much larger than \Hm. 
Such a behavior is reminiscent of that observed for the field-reinforced SC for $H \parallel b$ below \Hm\ \cite{RosuelPRX2023}, where the coupling strength increases on approaching \Hm. 
However, to date, a well-defined theoretical scenario for the field dependence of $\lambda$ in the hfSC phase is lacking.

We will discuss a proposal for such a model later in the paper.  
In order to determine minimal constraints from the data, we first analyze them without any field dependence of $\lambda$. 
We use the same strong-coupling model proposed for fields below \Hm in Ref.~\cite{RosuelPRX2023} in combination with the hypothesis that the paramagnetic limit is absent for the hfSC phase (as for a spin-triplet equal-spin-pairing (ESP) state). 
$\lambda$ is adjusted in order to have a large enough \Tc (in zero field) that could explain the survival of SC above \Hm. 
The orbital limit is mainly controlled by an (Fermi surface) averaged  renormalized Fermi velocity $\langle v_\mathrm{F}\rangle$, directed perpendicular to the magnetic field. 
The renormalization includes the effect of the pairing interactions. 
Hence, $\langle v_\mathrm{F}\rangle$ can be written as $\langle v_\mathrm{F}\rangle = \frac{\langle v_\mathrm{F}^{band}\rangle}{1+\lambda}$, where $\langle v_\mathrm{F}^{band}\rangle$ is a bare "band" averaged Fermi velocity (renormalized by all interactions but the pairing interaction), for which we used the same values along the $b$ and $c$ axis as in the low-field phase\ \cite{RosuelPRX2023} (see Supplementary Note~5 for more details on the model).
The required values of $\lambda$ range from 1.51 (at $25^{\circ}$) to 1.58 (at $35^{\circ}$) and the corresponding fits are shown in Fig.\ \ref{fig2}e. 

Remarkably, we could use the same $\langle v_\mathrm{F}^{band}\rangle$ as a control parameter of the orbital limit for the lfSC and hfSC phases at the same angle.
It seems to imply that correlations (except for the change of the value of $\lambda$) and the Fermi surface experience no dramatic change at \Hm. 
In other systems, where quantum-oscillation measurements could be performed, such as the well-documented case CeRu$_2$Si$_2$\ \cite{AokiPRL1993, FlouquetPhysicaB2002} as well as the uranium systems UPt$_{3}$\ \cite{JulianPRB1992} and UPd$_{2}$Al$_{3}$\ \cite{TerashimaPRB1997}, Fermi surface changes were observed across the metamagnetic field $H_m$, as well as heavy masses just above $H_m$. 
However, these heavy masses should be suppressed much faster by external magnetic field in cerium-based systems, which show a clearer trend to localization of the $f$-electrons under field and smaller Kondo temperatures than uranium systems. 
We will discuss later particular aspects of \UTe that explain why it preserves large effective masses above $H_m$, at least for the singular field orientations where SC reappears. 
From this first analysis, we conclude that the existence of the hfSC phase still requires an absent paramagnetic limit and large effective masses similar to the lfSC phase (same $\langle v_\mathrm{F}\rangle$). 
This, together with the enhanced (zero-field) critical temperature, is sufficient to explain that SC can survive at these record high fields. 
Our approach reproduces the overall temperature dependence of \Hc reasonably well and yields $\mu_0 H^\mathrm{max}_{\mathrm{c}2}\approx 73\,$T ($\pm 1\,$T) between $30$ and $35^\circ$. 
The obtained \Tc values, extrapolated to zero field, range between 3.2 and $3.6\,$K. 
In Fig.\ S4 we present a normalized comparison of $H_\mathrm{c2}(\theta)$ for both the hfSC and the lfSC phases. 
The remarkable anisotropy of \Hc in the hfSC with a peak around $\theta\approx 35^\circ$ is contrasted by the monotonically decreasing \Hc of the lfSC. 
The maximum \Hc value sets a record-breaking mark for SC emerging in a heavy-fermion compound to date.
The existence of heavy quasiparticles at fields above 40\,T means that renormalization of the effective masses by the Kondo effect is still effective above \Hm.

\subsection*{Strong suppression of the Hall effect in the vicinity of the hfSC phase}

\begin{figure*}[tbh]
	\centering
	\includegraphics[width=0.9\textwidth]{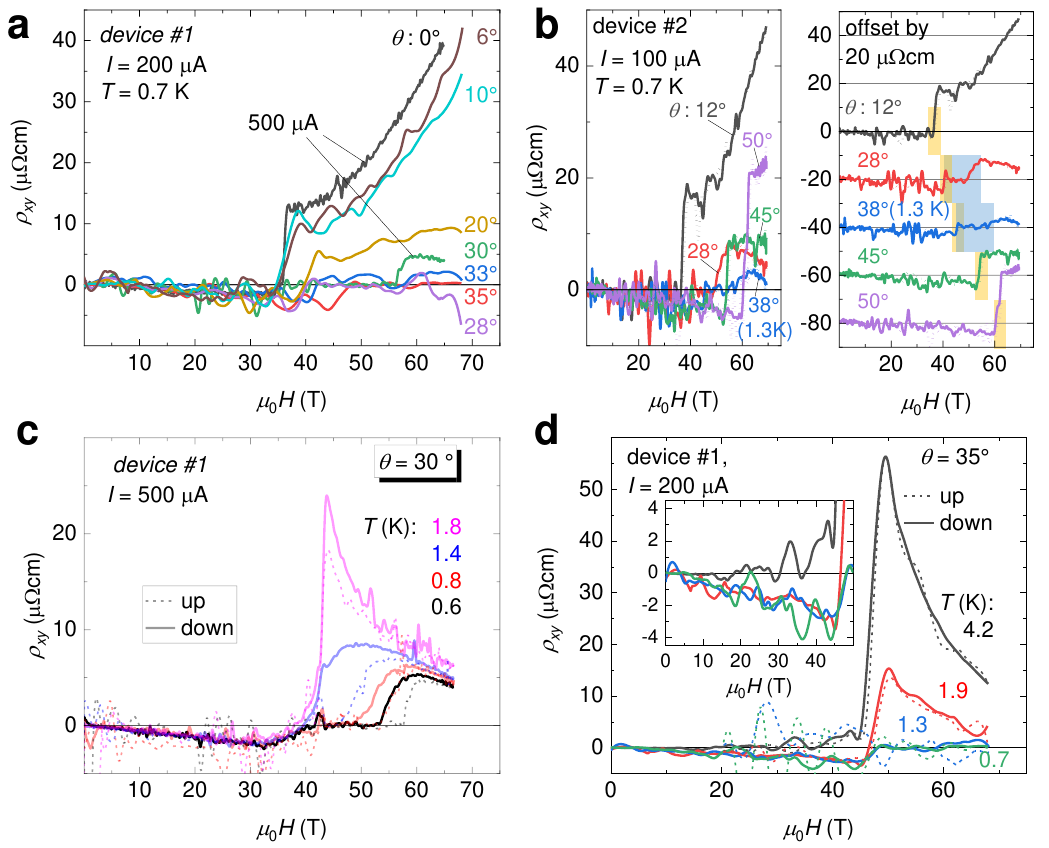}
	\caption{
    \textbf{High-field Hall effect of \UTe.} (a), (c), and (d) Hall resistivity of device $\#1$ recorded with two different currents for field-up and -down sweeps (c), (d) at fixed field orientations, $\theta=30^\circ$ and $35^\circ$, for various temperatures, and (a) at fixed temperature, $T=0.7\,$K, for various angles. 
    (b) Left panel: Hall resistivity recorded for device $\#2$ with $I=100\,\mu$A at fixed temperature and different angles. 
    Right panel: The curves were shifted by a constant offset for better visibility.
    \Hm and the hfSC region are highlighted by yellow and blue shaded bars, respectively. 
	Inset in (d): zoom into the region below 50\,T. 
	}
	\label{fig3}
\end{figure*}

In Figures\ \ref{fig3}a-d, we present Hall-resistivity data recorded in pulsed magnetic fields for devices $\#1$ and $\#2$ at two different currents and for various angles and temperatures. 
The Hall resistivity is composed of the ordinary component linked to the charge-carrier density and mobility, and an AHE component, whose origin is still the subject of intense research\ \cite{Nagaosa2010AHE}: 
it may have an intrinsic origin related to the topology of the electronic band structure, well identified in ferromagnets, or an extrinsic origin arising from different scattering mechanisms (skew-scattering or side-jump), all a consequence of spin-orbit interactions. 

In the case of heavy-fermion systems, even though there is no accepted complete microscopic theory\ \cite{Nagaosa2010AHE,Yang2013AHE}, the most successful interpretation of the AHE relies on skew scattering from local and itinerant $f$ electrons\ \cite{Fert1987Hall,Yang2013AHE}. 
An analysis of the electrical-transport coefficient obtained in steady fields up to 35\,T by Niu et al. pointed out that coherent skew scattering of the conduction electrons is the dominant contribution to Hall effect below about $20\,$K for $H\parallel b$\ \cite{NiuPRR2020}. 
We provide additional Hall data recorded at temperatures between 1.4 and 77\,K for device $\#3$ in the Supplementary Information (Fig.\ S2), consistent with the previous report\ \cite{NiuPRR2020}.

In the following, we will focus on the angular dependence of the Hall effect in \UTe. 
We recorded high-resolution Hall-effect data for two different transport devices with $I=500\,\mu A$, $200\,\mu$A, and $100\,\mu$A. 
The lower currents provide the least heating of the samples but a reduced signal-to-noise ratio.  
The overall low-field Hall signal acquires a negative magnitude and slope once the normal state is reached. 
At \Hm a sharp jump, similar to that of the magnetoresistivity, occurs and the Hall resistivity changes sign consistent with observations for $H \parallel b$\ \cite{NiuPRR2020}. 
We observe a drastic change of the high-field Hall signal with angle, see Fig\ \ref{fig3}a and b. 
The overall slope of $\rho_{xy}(H)$ changes from positive to negative at highest fields as we increase the tilt angle to about $28^\circ$. 
This is supported by the high-resolution (larger bias current) measurements of $\rho_{xy}(H)$ shown in Fig\ \ref{fig3}c and the higher-temperature measurements shown in Fig\ \ref{fig3}d. 
Moreover, the magnitude of $\rho_{xy}$ becomes strongly suppressed with a hardly discernible jump at \Hm$(68\,$T at $38^\circ)$. 
Interestingly, at $45^\circ$ and beyond, the feature at \Hm is visible again. 

Most importantly, at angles ranging from $28^\circ$ to $38^\circ$ the Hall resistivity is zero in the superconducting state just as the resistivity. 
This is most apparent in the high-resolution data recorded at $\theta=30^\circ$, with a current of $500\,\mu$A, shown in Fig.\ \ref{fig3}c. 
Even though there is a slight difference in the transition field between the up and down sweep, potentially originating from heating, all curves show a zero signal in the superconducting state.   
Previous pulsed-field resistivity and magnetization studies have already reported a zero-resistance state indicating the hfSC phase\ \cite{Ran2019extreme}. 
Nevertheless, a low resistivity may indicate a very metallic state, but may not be unambiguous proof for the presence of SC. 
Here, measurements of Hall resistivity can be of great help: They are sensitive to the nature and to the density of states near the Fermi level mainly responsible for the transport properties. 
This has been well demonstrated in layered delafossite compounds, where a super-low-resistive ground state was observed with a resistivity at 4.2\,K below $0.01\,\mu\Omega$cm (very hard to detect for bulk devices)\ \cite{HicksPRL2012,Kushwaha2015,Mackenzie2017}. 
In this particular case, a large mean free path reduces scattering, resulting in a hardly detectable resistivity response. 
Yet, the Hall resistivity remains non zero, signaling a well-established Fermi surface. 
Therefore, the vanishing of the Hall resistivity (within the noise) observed in the hfSC phase for \UTe, provides further proof for condensation of charge carriers in a superconducting state in the hfSC phase.
Note 1: At higher temperatures, $\rho_{xy}$ below \Hm gradually changes from negative to positive and crosses zero (see 4.2\,K data in the inset of Fig.\ \ref{fig3}d). 
Our resolution of a few micron thin device enables us to distinguish the weak negative low-field Hall effect from the zero signal in the hfSC region, best demonstrated in Fig.\ \ref{fig3}c. 
Note 2: in the low field regime, the resolution of $\rho_{xy}$ is degraded due to the small signal amplitude in the normal state. 
This, together with the increased heating at the end of the magnetic field pulses prevents the observation the effect of SC on $\rho_{xy}$ in the lfSC phase as clearly as in the hfSC phase. 

\begin{figure*}[tbh]
	\centering
	\includegraphics[width=1\textwidth]{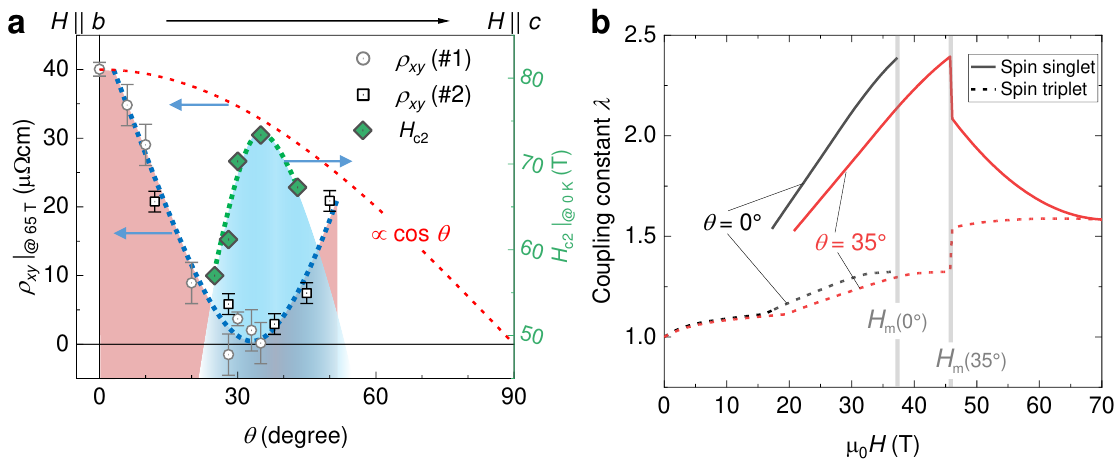}
	\caption{ 
    \textbf{Evidence for the Jaccarino-Peter compensation effect in \UTe.}
    \textbf{(a)} Angular dependence of the normal-state Hall resistivity at 65\,T and 0.7\,K of devices $\#1$ and $\#2$, taken from Fig.~\ref{fig3}a and b. 
	The blue dashed line is a guide to the eye that highlights the observed strong suppression around $\theta\approx 30-35^\circ$. Green diamonds are \Hc values extrapolated to zero temperature, taken from Fig.\ \ref{fig2}e.
    The red dashed line follows $\cos{\theta}$.
    \textbf{(b)} Strng-coupling constant $\lambda$ used for describing the experimental $H_\mathrm{c2}(T)$ data points presented in Fig.~\ref{fig2}e, 
    within the two proposed scenarios: 
    Solid (dashed) lines are with (without) paramagnetic limitation of \Hc, assuming a spin-singlet (spin-triplet-ESP) state.
    Below \Hm at $0^\circ$ ($H \parallel b$), the curves are those from Ref.\ \cite{RosuelPRX2023}.
    Below \Hm at $35^\circ$, the curves are scaled by $1/\cos(\theta)$, assuming that only the $b$ component $H$ determines $\lambda(H)$.
    The field dependence of $\lambda$ above \Hm is based on the compensation mechanism in both cases as explained in the main text and in Supplementary Note\ 5. The spin-singlet result resembles the $H$ dependence reported for the specific heat (for $H \parallel b$) or the $A$ coefficient around \Hm\ \cite{Miyake2019, Imajo2019, RosuelPRX2023, KnafoJPSJ2019}. 
    }
	\label{fig4}
\end{figure*}

The steep angular suppression of the high-field Hall effect signal is shown in Fig.\ \ref{fig4}. Therein, we plot the $\rho_{xy}$ at a fixed magnetic field of $68\,$T against the tilt angle. 
For angles above approximately $40^\circ$, $\rho_{xy}$ in the normal state recovers again and reaches values close to those expected from a conventional $\cos{\theta}$ scaling behavior, indicated by the red-dashed line.
The mechanism behind the drop in $\rho_{xy}(\theta)$ is a puzzle, particularly when compared to previous work that explored the electronic properties of \UTe at field orientations around $30^\circ$ tilt within the $(b,c)$ plane.
Indeed, previous magnetization measurements observed no significant change of the magnetization jump at \Hm for fields along the $b$ axis and around $30^\circ$\ \cite{Ran2019extreme,MiyakeA2022}. 
Similarly, the resistivity does not show significant changes around \Hm for both field orientations\ \cite{KnafoCommPhys2021}, indicating that neither the elastic nor the inelastic scattering display a considerable evolution with angle. 
Therefore, we expect the AHE component, which is directly proportional to $M$ and to the resistivity or the square of the resistivity, to remain (roughly) constant with angle. 

Recent dHvA studies, which confirmed the Fermi-surface topology predicted by band-structure calculations\ \cite{Aoki2022dHvA,Eaton2023}, may hint at specific properties linked to the $\theta\approx30^\circ$ field orientation.
In particular, the warping of the cylindrical Fermi surfaces could meet the so-called “Yamaji magic-angle” condition\ \cite{Yamaji1989} that can induce a suppressed conduction for a particular field orientation and, thus, affects the density of states at the Fermi edge. 
To date, the exact Fermi-surface topology in the high-field regime above \Hm has not been revealed.
Thus, the potential influence of Fermi-surface anomalies on the Hall coefficient above \Hm is unknown.

\subsection*{Analysis of the Hall effect and connection with the hfSC phase}

In the following, we propose a scenario for the reentrant hfSC phase, supported by the analyses of our high-field torque and Hall-effect results. 
We will show that the origin of the Hall effect above \Hm should be revisited, arising most likely from an intrinsic topological contribution. 
Hence, suppression of the Hall effect is best explained by that of the band polarization, leading naturally to a Jaccarino-Peter mechanism for the hfSC phase. 

In the Supplementary Information (Fig.\ S5), we present an analysis of the Hall data along the lines of Niu et al.\ \cite{NiuPRR2020}. 
Under the assumption that skew scattering (directly proportional to the product $\rho_{xx}^2 M$) is the dominating extrinsic component at low temperatures and with the inclusion of already published magnetization data\ \cite{Ran2019extreme,Miyake2019}, we can extract the normal (orbital) Hall coefficient $R_\mathrm{H}$ at $\theta=30^\circ$ from the intercepts in Fig.\ S5 (second and third column). 
Apparently, $R_\mathrm{H}$ jumps by a factor of two, i.e., $0.05\,\Omega\mathrm{cm}/T\rightarrow 0.1\,\Omega\mathrm{cm}/T$ when transitioning from below to above \Hm. 
Intriguingly, the high-field value is almost one order of magnitude smaller than what was reported for the $H\parallel b$ orientation\ \cite{NiuPRL2020}. 
This analysis implies, however, that the proposed\ \cite{NiuPRL2020} strong suppression (by a factor 10) of the charge-carrier density for $H\parallel b$ is not present anymore for tilted field.  
Moreover, significant changes in the charge-carrier density at \Hm have not been confirmed by any other reported quantity, such as the specific heat\ \cite{Miyake2019, Imajo2019, RosuelPRX2023} or the $A$ coefficient of the resistivity\ \cite{KnafoCommPhys2021}. 
A dramatic suppression of the density of states would also be hard to reconcile (if persistent for tilted fields) with the appearance of the hfSC phase. 
We, therefore, argue that the conventional interpretation in terms of normal and skew-scattering dominated contributions to the Hall effect proposed in Ref.\ \cite{NiuPRR2020} does not hold in the case of \UTe. 
Indeed, the general understanding of mechanisms behind the AHE has significantly improved in recent years\ \cite{Onoda2006,Nagaosa2010AHE}. 
In particular,the role of intrinsic (topological) contributions, expected to scale with $\rho_{xx}^2 M$, has been discussed\ \cite{Onoda2006,Nagaosa2010AHE}. 
Such contributions depend on topological invariants associated with the band polarization. 
They are already present in zero field for ferromagnetic systems.
The band structure of \UTe may host topological features such as Weyl nodes near the Fermi edge\ \cite{Shishidou2021TologigicalBand}. 
Such contributions should dominate the AHE when the resistivity is in the range between 1 and 100\,$\mu \Omega cm$\ \cite{Nagaosa2010AHE}. 
Moreover, the dependence on magnetization in ferromagnets arises not from magnetic interactions, but simply from the domain alignments: 
In other words, if such a contribution appeared above \Hm due to a sudden band polarization at the metamagnetic transition, it would keep a $\rho_{xx}^2$ dependence, but the $M$ factor might be meaningless. 
Hence, the strong negative drop of the normal Hall coefficient reported by Ref.\ \cite{NiuPRR2020} can also be explained by the emergence of a strong intrinsic anomalous Hall effect (iAHE) at \Hm.
Furthermore, the role of skew scattering could have been largely overestimated.
As a consequence, our observed angle-dependent suppression of the Hall effect around $\theta\approx30^\circ$ in the $(b,c)$ plane should then reflect the suppression of this iAHE. 
With an almost angle-independent jump in the magnetization at \Hm (at least within the angular range, where hfSC exists\ \cite{Ran2019extreme, Miyake2019}), the steep decline of the iAHE contribution suggests a suppressed influence of the topological aspect in the band structure on the AHE. 

Band splitting with avoided level crossing is key for this intrinsic contribution to the Hall effect\ \cite{Nagaosa2010AHE}. 
So an appealing possibility is that the suppression of the iAHE contribution arises from a strong decrease of the band polarization in this angular range.
It could result from a compensation between the applied field and an exchange field between the conduction bands and local magnetic moments, polarized by the metamagnetic transition.
The background picture for this scenario is that a main contribution to the magnetization of \UTe arises from localized $5f$-electrons. 
This is consistent with the large nearest-neighbor distance, far exceeding the Hill limit\ \cite{Ikeda2006}.
It is furthermore supported by band-structure calculations that predict a Fermi surface dominated by Te-$5p$ and U-$6d$ electrons (partly hybridized with U-$5f$), with at most only small $5f$-electron pockets\ \cite{IshizukaPRB2021}. These, however, have not been observed by experiments to date\ \cite{Aoki2022dHvA}. 
In such a scheme, the jump of the magnetization at \Hm arises mainly from local moments having (antiferromagnetic) exchange coupling with the conduction bands, a very natural scheme for a Kondo system. 
A reduction of the band polarization, arising from the compensation between exchange and applied field above \Hm also explains why we can fit \Hc in the hfSC phase with the assumption of unaltered $\langle v_\mathrm{F}^{band}\rangle$ values as compared to the lfSC phase (Fig.\ \ref{fig2}e.): 
the main effects of the metamagnetic transition on the Fermi surface then disappear. 
More importantly, this compensation between $H$ and the ``molecular'' exchange field is  instrumental for the so-called Jaccarino-Peter mechanism\ \cite{Jaccarino1962,Meul1984,Uji2001} that could account for the reentrant hfSC phase. 

\subsection*{Jaccarino-Peter compensation effect in \texorpdfstring{\UTe}{UTe2}}

Before we discuss this Jaccarino-Peter (JP) mechanism, let us summarize briefly the present situation for the various superconducting phases in \UTe at ambient pressure.  
In zero field, there is a consensus for \UTe being recognized as a candidate spin triplet superconductor with a B$_{3u}$ or A$_u$ symmetry.  
Finer details such as the nodal structure are still under debate\ \cite{AokiJPCM2022}. 
Recently, several experiments have revealed a clear phase transition to another superconducting phase for field along the $b$ axis above $\approx15\,$T at low temperatures\ \cite{RosuelPRX2023, SakaiPRL2023, KinjoPRB2023}. 
Theoretical proposals anticipated a transition to a B$_{2u}$ symmetry ($d$ vector with no component along the $b$ axis).
In contrast, thermodynamic experiments have revealed drastic changes between the two phases, suggesting a change of the pairing mechanism in addition to a symmetry change\ \cite{RosuelPRX2023}. 
It has also been shown that a spin-singlet state can account for the observed strong broadening of the superconducting anomaly as well as its angular dependence\ \cite{RosuelPRX2023}.
Regarding the hfSC phase, as opposed to the phases below \Hm, there are no theoretical models yet, and the common wisdom is that it should be spin-triplet to survive such high fields. 
The initial proposed mechanisms were a Jaccarino-Peter or a Lebed mechanism\ \cite{Ran2019extreme}, dismissed or abandoned for that of a "Landau level superconductivity"\ \cite{Ran2021expansion, Frank2023orphan}. 
This last hypothesis is rather surprising when contrasted with the existence of the hfSC phase even in very dirty systems\ \cite{Frank2023orphan}. 

In the following, we propose a Jaccarino Peter scenario mainly relying on a spin-singlet state. 
As shown already for the field-reinforced superconducting phase along the $b$ axis, spin-singlet SC is able to survive in such high fields supported by strong-coupling effects and the field-induced reinforcement of the pairing strength\ \cite{RosuelPRX2023}. 
At the root of the JP mechanism is a compensation between the external field, $H$, and an internal exchange field, \He\ \cite{Jaccarino1962}.
The latter is associated with the polarization of local magnetic ions and acts on the spin of the itinerant quasiparticles.
Compensation is possible only if \He is opposite to $H$, which requires an antiferromagnetic exchange coupling between local and itinerant spins.
In the case of \UTe, the local moments originate from the uranium ions. 
At the mean-field level, \He can be expressed as: \He$=J_c<M_c>\hat{c} + J_b<M_b>\hat{b}$, with $J_c$ and $J_b$ and $<M_c>$ and $<M_b>$, respectively, the anisotropic exchange constants and magnetization components along the $c$ and $b$ axes. 
Hence, at finite tilts within the $(b,c)$ plane, the direction of \He is most likely not perfectly collinear with $H$. 
Nevertheless, antiferromagnetic coupling (negative $J_c$ and $J_b$) is quite natural for such a Kondo-lattice system.
If $H$ and \He compensate each other, then the itinerant quasiparticles feel no Zeeman field and they should lose their polarization: 
Our Hall-effect results suggest that in the angular range around $30^\circ$, the compensation between both fields is quite efficient, at least around $70$\,T. 
The marked decrease of the torque at $25^\circ$ and low temperatures, mentioned earlier, also indicates that for this angle, the magnetization above \Hm is closer to the field direction.
This is beneficial for the compensation of \He by $H$ in neighboring angles (once taking into account exchange anisotropy). 
We see two possible ways for how the hfSC phase arises via this JP compensation mechanism. 

In the first scenario, the superconducting pairing is restored by an absence of band polarization (and at the opposite, suppressed when the magnetization is saturated).
This would work both for a spin-singlet and a spin-triplet superconducting order parameter.

In the second scenario, \He directly counters the paramagnetic limit enabling the restoration of SC:  
this would be a ``true'' JP compensation\ \cite{Jaccarino1962, FischerHPA1972, OdermattHPA1981}, requiring that the hfSC phase is spin singlet or spin triplet with a sizeable $d$ component along the applied field: 
\He solely acts on the spins, i.e., when the compensation of $H$ and \He becomes perfect at a certain field, 
\Hc remains restricted only by the orbital limit. 
Stunning examples were found among organic superconductors with field-reentrant phases attributed to the JP compensation effect\ \cite{KonoikePRB2004,UjiJPSJ2006}.
However, the observed angular dependencies were extremely sensitive to the field alignment due the huge anisotropy of the orbital limit in these 2D materials. 
In both proposed scenarios, a natural assumption is that the reentrant hfSC phase is a resurgence of the field-reinforced superconducting phase observed for a narrow angular range (few degrees) about $H \parallel b$, persistent up to \Hm.
Otherwise, yet another pairing mechanism should take place in \UTe above \Hm. 

We have discussed already (see Fig.~\ref{fig2}e) how the hfSC phase could exist for \Hc only limited by the orbital effect, thanks to an increase of \Tc of up to about 3\,K without any change of the bare $v_\mathrm{F}$ as compared to the lfSC phase. 
This hypothesis of pure orbital limiting requires an ESP spin-triplet state for the hfSC phase.
Hence, the compensation effect could only act on the value of $\lambda$ (first scenario).
The results of Fig.~\ref{fig2}e show that in such a case, $\lambda$ (see dashed lines in Fig.\ \ref{fig4}b) would be essentially field independent above \Hm with a value changing only little (between 1.5 and 1.6) within the angular range of the hfSC phase.
As a consequence, $\lambda$ above \Hm should grow with the tilt angle.
This, however, stands in contrast to the quick vanishing of the field-reinforced SC beyond only few degrees tilt. 
Moreover, for the JP compensation to work effectively, $\lambda$ at finite angle should be at most of the order of that along the $b$ axis in the field-reinforced phase just below \Hm.

By contrast, the second scenario involving a true JP effect does not suffer from these caveats, if we assume a spin-singlet phase below \Hm for $H \parallel b$\ \cite{RosuelPRX2023}. 
The distinction to the first scenario is that in this case below \Hm, SC is mainly controlled by the Pauli depairing. 
Hence, the fast suppression of SC for only small tilts away from $H\parallel b$ can be attributed to the lowering of the Pauli limit at constant field, when $\lambda(H)$ decreases due to the increase of \Hm with angle (see Supplementary Note~5).
Then, even partial compensation of this paramagnetic limit by \He above \Hm at finite angle can restore SC without requiring to surpass $\lambda$ for $ H \parallel b$. 

In order to obtain a proper model that can describe the location of the hfSC pocket in the $(\theta,T,H)$ phase space, and notably the $T$ dependence of \Hc at a given $\theta$ in the JP scenario, we need to know the $H$ dependence of $\lambda$ together with the degree of compensation of the paramagnetic limit.
Presently, too little is known about the magnetization and \He above \Hm (see discussion in Supplementary Note~5). 
Thus, there are (too) many possible tuning parameters for a comprehensive quantitative model of the hfSC pocket.
Nevertheless, we can attempt a modelling of our angle-dependent \Hc results. 
In order to fix the $H$-dependent compensation of the paramagnetic limit controlled by $g\mu_\mathrm{B}(H - H_\mathrm{ex})$, we use \He$=70\,$T under the assumption that it remains constant and parallel to $H$. 
Hence, $\lambda(H)$ can be extracted from the data of Fig.~\ref{fig2}e. 
The result is shown in Fig.~\ref{fig4}b. 
We find that the JP compensation scenario leads to a decrease of $\lambda$ just above \Hm, diminishing further for $H>H_\mathrm{m}$. 
Again, the key point here is the dominant role of the paramagnetic limit, i.e., this controls the disappearance of SC at finite angle below \Hm and its reentrance above \Hm due to the compensation. 
More details on the model are given in Supplementary Note~5. 
In addition, the proposed mechanism can also explain the recently reported pressure dependence of the hfSC phase\ \cite{Ran2021expansion}. 
Under pressure, SC was found to survive at finite angle up to \Hm, or even to exist above \Hm detached from the metamagnetic transition line.

The mechanism behind the hfSC phase and its relation to the SC at lower fields is under hot debate\ \cite{Ran2019extreme,Frank2023orphan} and still without even a qualitative satisfying scenario. 
Here, we show how the JP compensation effect, dismissed by previous studies, can explain the hfSC phase. 
This is supported by the vanishing of the Hall effect. 
This scenario of spin-singlet SC in the hfSC phase is also connected to the same state proposed for the field-reinforced superconducting phase emerging below the metamagnetic transition for $H \parallel b$\ \cite{RosuelPRX2023}. 

In summary, our study features insights on the enigmatic high-field properties of the putative spin-triplet heavy-fermion superconductor \UTe.
We demonstrate by torque magnetometry that the magnetization jump at the metamagnetic transition and the applied field are noncollinear, keeping a large component along the $b$ axis.
This is probably related to the observed $1/\cos\theta$ dependence of the metamagnetic transition field, \Hm. 
We studied angle-dependent magnetotransport in 70\,T pulsed magnetic fields. 
Here, we focused particularly on the distinct high-field superconducting phase induced just above \Hm for tilt angles of around $35^\circ$ within the $(b,c)$ plane surviving very high field values above $40\,$T. 
We have determined the angular-dependence of the upper critical field, \Hc, in this phase, reaching a maximum of $\mu_0$\Hc$\approx 73\,$T. 
This value is amongst the highest reported for heavy-fermion superconductors.
Our studies reveal an apparent correlation between the upper critical field, \Hc, and the normal-state Hall effect at very high fields. 
The latter exhibits a minimum with an almost complete suppression, precisely where the reentrant hfSC emerges and reaches its maximum robustness. 
The analyses of this correlation hints at a compensation mechanism as the potential origin of both phenomena: 
In the angular region around $35^\circ$, compensation between the exchange field above \Hm and the applied field, such that band polarization is strongly reduced, is consistent with our observations. 
A reduced band polarization can lead to the suppression of the dominant AHE contribution and to a reentrant superconducting phase (Jaccarino-Peter effect). 
Our results provide a guide for future experiments and theory that will show more quantitatively if and how this may appear. 
Such a scenario puts specific constraints on the potential order parameter of the superconducting phase discussed in our work.
Solving the riddle of how Cooper pairs, built by heavy quasiparticles, can survive in extreme magnetic fields will certainly help advance our fundamental understanding of unconventional superconductors.

\section*{Methods}

\textbf{Crystal growth:} The \UTe single crystals were prepared as described in Ref.\ \cite{AokiJPSJ2019}. 
All single crystals were prepared by the chemical vapor transport method with iodine as transport medium. A starting ratio of U:Te $=2:3$ has been used, and the quartz ampules were heated slowly up to a final temperature of $1000^\circ$C on one side and $1060^\circ$C on the other side and this temperature gradient was maintained for 18 days. The ampules were slowly cooled down to ambient temperature during 70 hours.

\textbf{Microcantilever torque magnetometry:} 
For magnetic-torque experiments, we cut samples with dimensions ($100\times 20\times 3)\,\mu$m$^3$ from a single crystal using focused-ion-beam (FIB) assisted etching. 
We used a Wheatstone-bridge-balanced piezo-resistive cantilever (eigenfrequency $\sim 300\,$kHz)\ \cite{Ohmichi2002}. 
The sample was attached by Apiezon (N) grease. 
The setup was mounted on a rotator, such that the angle between field and cantilever could be varied, and installed in a $^3$He cryostat. Pulsed magnetic fields of up to $70\,$T were applied. An example picture of the microcantilever including a sample attached to it is presented in the inset of Fig.\ \ref{fig1}a.

\textbf{FIB-microfabrication of transport devices:} 
Device $\#1$, shown in Fig.\ \ref{fig1}b, was fabricated in the following steps: 
First, a slice $(150\times 20\times 2)\mu\mathrm{m}^3$ was separated out of the crystal using FIB and transferred \textit{ex situ} onto a sapphire chip. 
Next, an approximately 150\,nm thick layer of gold was sputter deposited covering a rectangular area around and including the sample slice. 
In a next step, carbon-rich platinum was deposited in a FIB system at the two ends and at six side points around the sample slice (see Fig.\ \ref{fig1}b). 
The platinum fixations establish a galvanic connection between the gold layer on the chip and the top surface of the sample. 
Next, the gold layer was partially etched away from the central top surface of the sample by ions. 
Then, a focused ion beam was applied to cut trenches into the gold layer and the sample in order to create well-defined terminals. 
Resistances of a few ohms were achieved. 
In the end, a droplet of transparent unfilled Stycast hardened in vacuum was used to protect the structured device from air. 
We fabricated three different devices for this study with the following width, thickness, and length ($w\times d\times l$) between the contacts: device $\#1$ $(10\times2.7\times75)\,\mu$m$^3$; device $\#2$ $(4\times2.9\times58)\,\mu$m$^3$; device $\#3$ $(7.1\times4.5\times48.5)\,\mu$m$^3$. 

\textbf{Magnetotransport measurements:} 
We performed steady-field characterization measurements in an Oxford dilution refrigerator equipped with an 18\,T superconducting magnet. 
We measured the resistance with a standard a.c. four-point lock-in technique. 
We conducted pulsed high-field experiments at the Dresden High Magnetic Field Laboratory in a $60\,$T and $70\,$T pulsed-magnet systems with a pulse duration of 25\,ms and 150\,ms, respectively, equipped with either $^4$He and $^3$He cryostat inserts.


\section*{References}
\bibliographystyle{naturemag}
\bibliography{UTe2literature}

\section*{Acknowledgements}
We thank J. Flouquet, H. Suderow, A. Miyake, Y. Yanase, M. Zhitomirsky and J. Gayles for fruitful discussions. 
The help by N. Nagaosa in the analysis of the anomalous Hall effect was particularly enlightening. 
We thank A. Mackenzie for continuous support. 
We thank S. Seifert for his technical assistance. DA acknowledges support by KAKENHI(JP19K03736, JP19H00646, JP20K20889, JP20H00130, JP20KK0061, JP22H04933), ICC-IMR. 
We acknowlege the support from GP-spin at Tohoku University and JSPS. AP, GL, GK and JPB acknowledge support from CEA Exploratory program TOPOHALL, the French National Agency for Research ANR within the project FRESCO No. ANR-20-CE30-0020, FETTOM No. ANR-19- CE30-0037 and SCATE ANR-22-CE30-0040. 
TH acknowledges support from the German Research Foundation (DFG) Grant No. HE 8556/3-1. 
We acknowledge the support of the HLD at HZDR, member of the European Magnetic Field Laboratory (EMFL), and the DFG through the W\"urzburg-Dresden Cluster of Excellence on Complexity and Topology in Quantum Matter - ct.qmat (EXC 2147, project-id 390858490).

\*section{Author contributions}
T.H., K.S., M.Ki., and M.K\"o. were involved in the preparation of the devices by FIB microfabrication. 
T.H., K.S., A.M., T.F., and M.Ki. performed the pulsed-field experiments. 
T.F. and T.H. performed high-field torque magnetometry. 
J.H., J.S., and T.H. performed low-field characterizations.
G.L., D.A., and G.K. prepared the high-quality samples. 
J.-P.B., T.H., K.S., and A.P. analyzed the upper critical field and the Hall-effect data. 
T.H., M.Ki., K.S., T.F., J.H., M.K\"o., I.S., A.P., D.A., G.K., J.W., and J.-P.B. took part in discussing the results and writing the manuscript. 

\newpage
\setcounter{figure}{0}
\makeatletter 
\renewcommand{\thefigure}{S\@arabic\c@figure}
\makeatother

\section*{Supplementary Information}

\subsection*{Supplementary Note 1: Torque magnetometry results}

\begin{figure}[tbh]
	\centering
	\includegraphics[width=.85\textwidth]{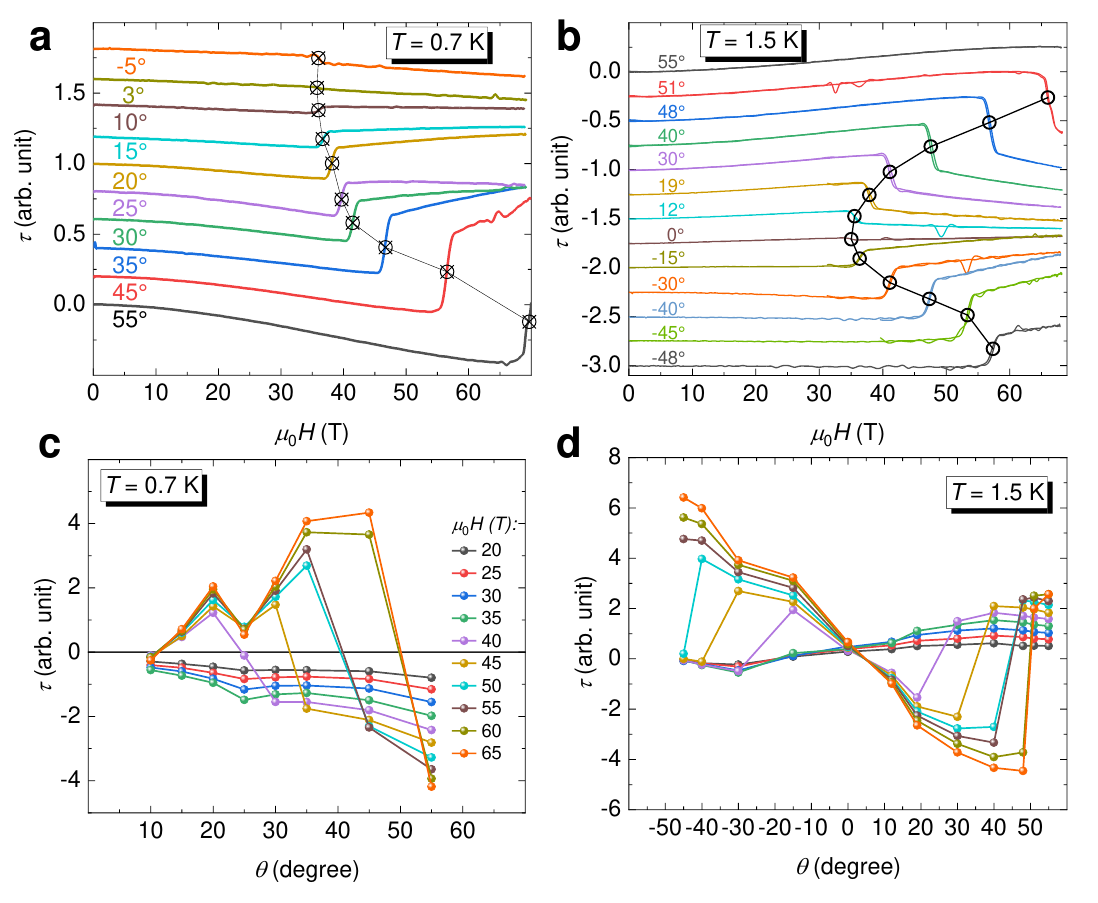}
	\caption{\textbf{Magnetic torque of \UTe.} (a), (b) Magnetic torque recorded at 0.7\,K and 1.5\,K for various angles in the $(b,c)$ plane. The tilt angle $\theta=0^\circ$ corresponds to $H\parallel b$. (c), (d) Amplitude of the torque determined at fixed field values plotted against the angular tilt. Note: The color-code for the field values is the same in both figures. The asymmetry between positive- and negative-angle curves is caused by the asymmetric deflection stiffness of the piezoresistive microcantilever.
    }
	
	\label{figS-Torque}
\end{figure}

In Fig.\ \ref{figS-Torque}, we present the torque data recorded at 0.7 and 1.5\,K in pulsed magnetic fields up to 70\,T. 
We applied a constant offset to the curves at different fixed angles for better visibility. 
The transition-field values, presented in the main text in Fig\ 1a, are marked at half the height of the jump in $\tau$. 
In figures (c) and (d) we show the same torque data plotted against the tilt angle $\theta$, for selected (marked) field values between 20 and 65\,T in 5\,T steps.

\newpage

\subsection*{Supplementary Note 2: Basic transport characterization}

\begin{figure}[b]
	\centering
	\includegraphics[width=0.9\textwidth]{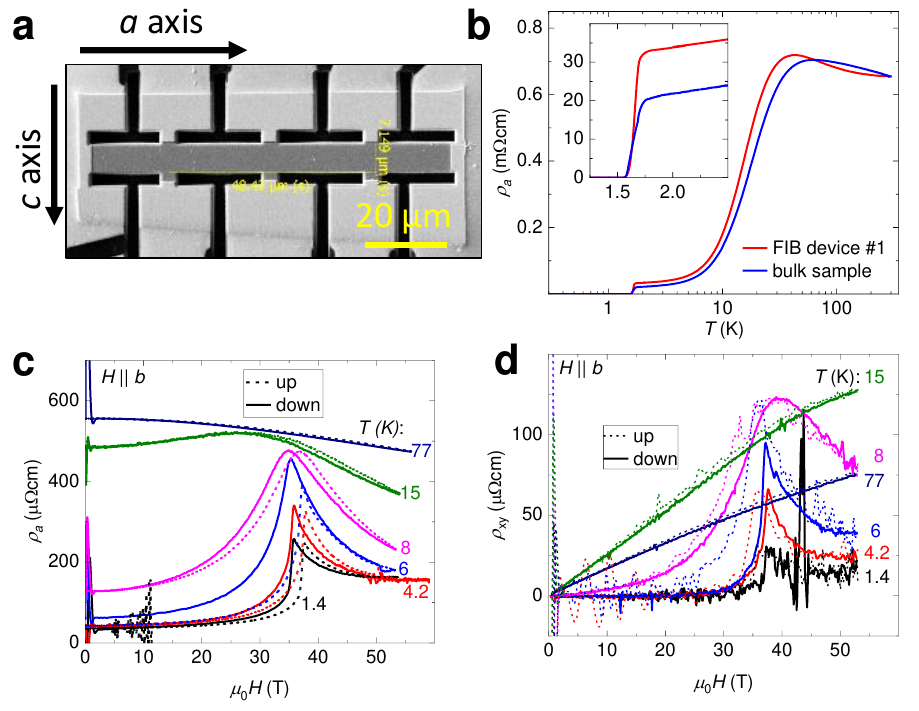}
	\caption{\textbf{Basic transport characterization of \UTe FIB devices.} 
    (a) SEM image of Hall-bar device $\#3$.
    (b) Comparison of zero-field $a$-axis resistivity for a bulk Hall-bar device and the FIB-microstructured device $\#1$, with dimension of $(2\times0.5\times0.1)\,$mm$^3$ and $(13\times5\times2)\,\mu\mathrm{m}^3$, respectively. The inset highlights the matching \Tc. 
    (c) Magnetoresistivity and (d) Hall resistivity of device $\#3$ recorded in a 60\,T short-pulse (25\,ms) magnet at various temperatures. 
    These results match nicely with data published previously for bulk samples~\cite{KnafoJPSJ2019}, confirming that FIB treatment does not alter the transport properties.}
    \label{figS-MR-D3}
\end{figure}
The FIB microfabrication process introduces an amorphized layer enriched with implanted Ga atoms (thickness: $\sim 20\,$nm). 
Hence, the standard first step in characterization of a FIB-cut (An example image is shown in Fig.\ \ref{figS-MR-D3}a) device is to test the zero-field resistivity. 
In Fig.\ \ref{figS-MR-D3}b 
we present a comparison of the $a$-axis resistivity for the bulk single crystal and device $\#1$.
Apparently the overall residual-resistance ratio and \Tc has not been altered by the fabrication procedure. Slight deviations may originate from differences in the homogeneity of stress induced by the sample substrate. 

Figure\ \ref{figS-MR-D3}a shows an scanning-electron-microscope (SEM) image of device $\#3$.
The measured $\rho_{a}(H)$ and $\rho_{xy}(H)$ reproduce results previously reported for bulk samples to temperatures of up to 77\,K (Figs.\ \ref{figS-MR-D3}c and d)\ \cite{KnafoJPSJ2019}. 
The measurements for this device were carried out in a pulsed magnet with a shorter pulse duration ($\sim 25\,$ms). 
This is why there is a significant hysteresis between the up- (dotted) and the down-sweep (solid) curves. 
Device $\#3$, however, did not survive transfer in air to perform additional experiments. 
The air sensitivity of \UTe may be responsible for a fast deterioration of the contacts. 
Consequently, the devices labeled with $\#1$ and $\#2$ were sealed by enclosing them with epoxy, for further details see the Methods section.
\newpage

\subsection*{Supplementary Note 3: Additional high-field magnetoresistivity data for device \texorpdfstring{$\#2$}{number 2}}
\begin{figure}[tbh]
	\centering
	\includegraphics[width=0.9\textwidth]{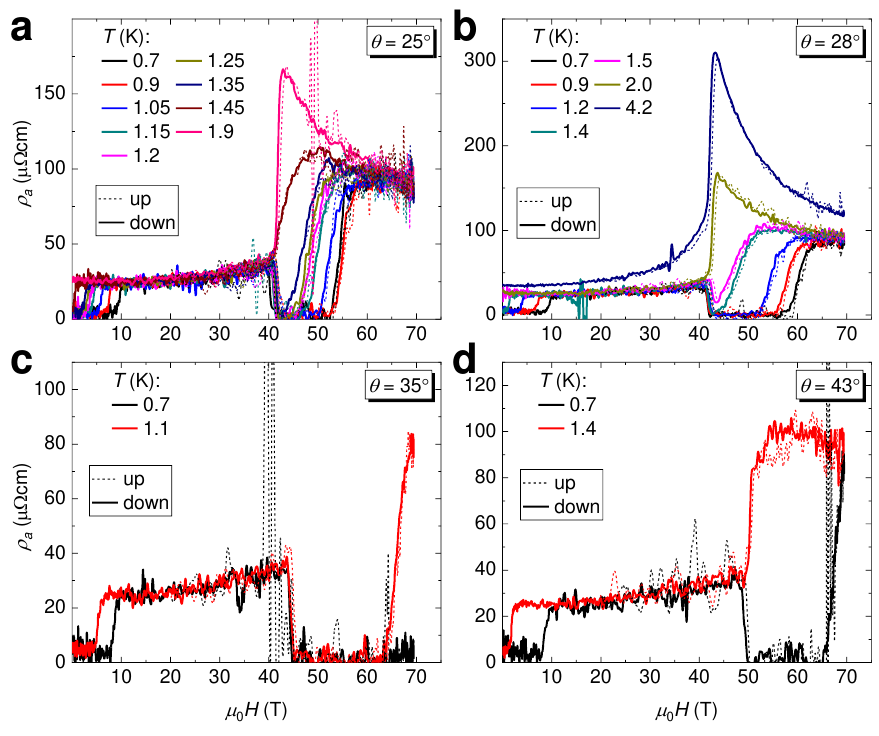}
	\caption{\textbf{Magnetoresistivity of \UTe recorded for device $\#2$ at four fixed angles} with applied current of $100\,\mu$A in a 70\,T pulsed magnet. Up and down seeps are marked by dotted and solid lines, respectively.
	}
	\label{figS-MR-D2}
\end{figure}

In Fig.\ \ref{figS-MR-D2} we provide data recorded for device $\#2$ at various angles and temperatures. The critical-field values presented in Fig.\ 3c and d were extracted from this data set.
\newpage

\subsubsection*{Anisotropy of \texorpdfstring{\Hc}{Hc2} in the high- and low-field superconducting phases}
\begin{figure}[tbh]
	\centering
	\includegraphics[width=0.55\textwidth]{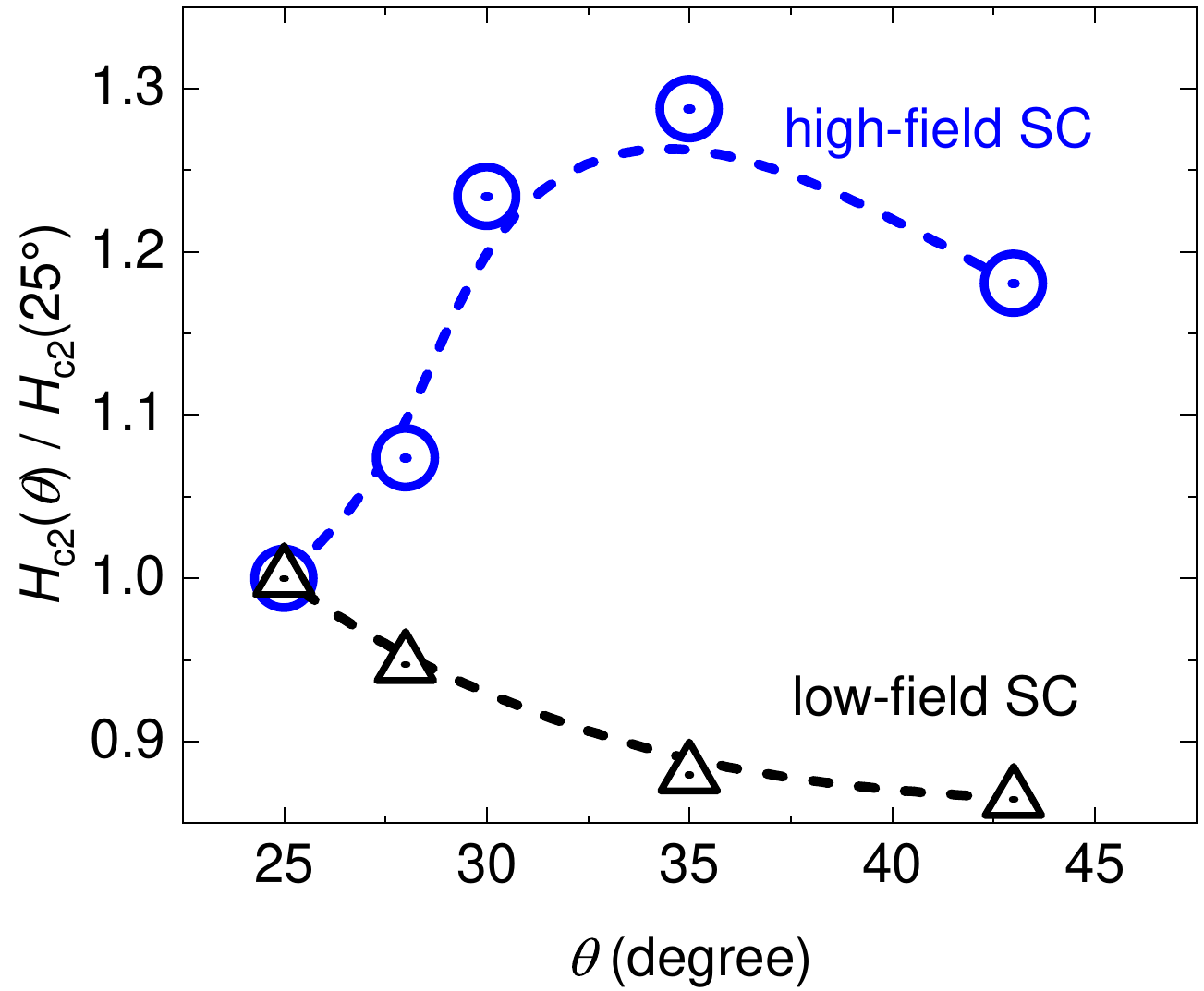}
	\caption{\textbf{Anisotropy of \Hc in the hfSC and lfSC phases.} Normalized upper-critical-field values at absolute zero in the hfSC and lfSC phases extracted from in Fig.\ 2e. The data were normalized to the respective \Hc values at $25^\circ$. Dashed lines are guides to the eye. 
	}
	\label{figS-Hc2Aniso}
\end{figure}
\newpage

\subsection*{Supplementary Note 4: Anomalous Hall effect at tilt angle}

\begin{figure}[tb]
	\centering
	\includegraphics[width=0.85\textwidth]{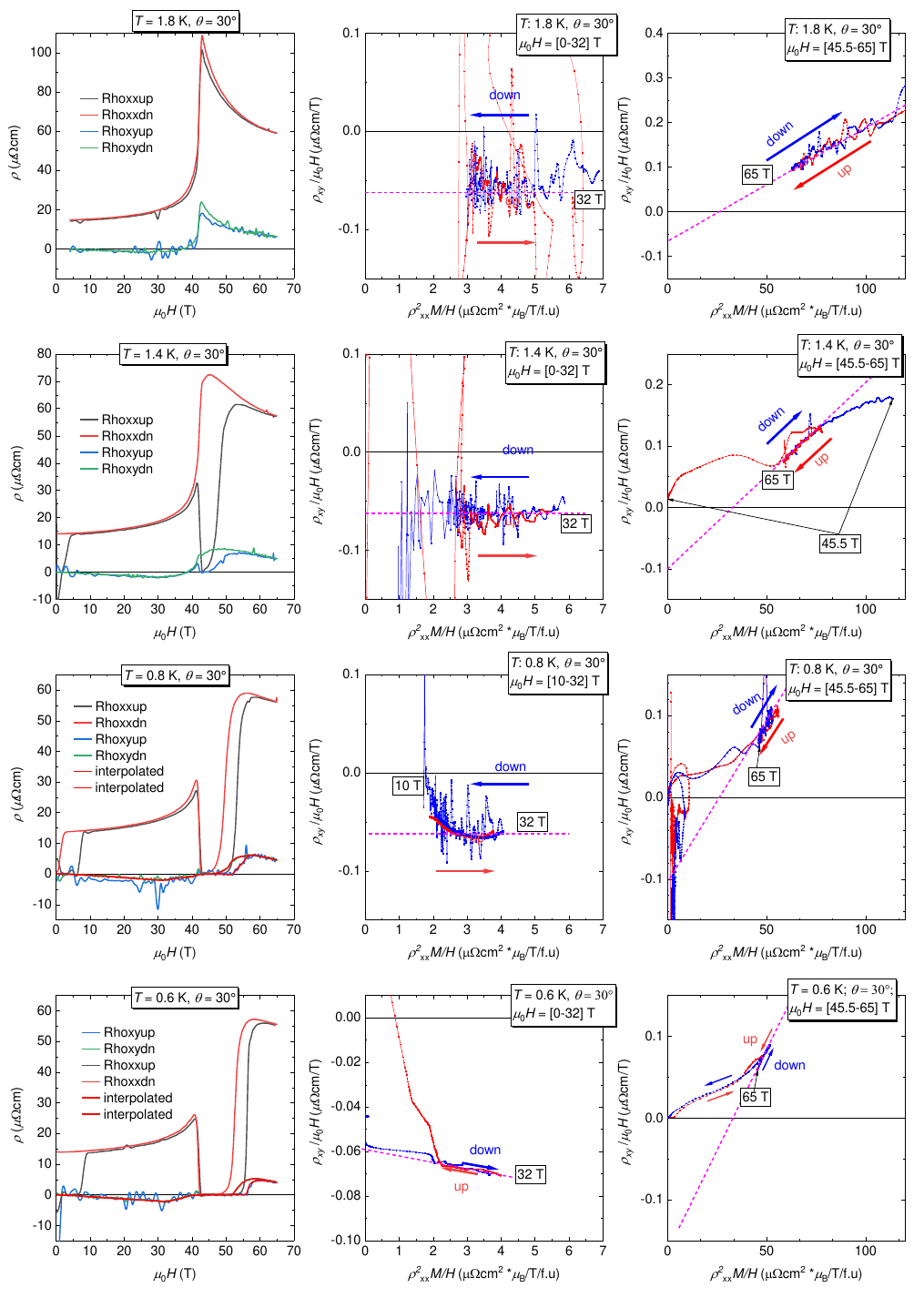}
	\caption{\textbf{Analysis of the AHE at $\theta=30^\circ$ following Refs.\ \cite{Yamada1993,NiuPRR2020}.} Left column: data set recorded for device $\#1$ at $\theta=30^\circ$ with $I=500\,\mu$A for $T=0.6$, $0.8$, $1.4$, and $1.8\,$K, respectively. Middle and right column: $\rho_{xy}/H$ against $\rho_{xx}^2M/H$ in the low and high-field range, respectively. Note: Dashed magenta lines are linear fits used to extract $R_0$ from the intercepts. }
	\label{figS-AHE}
\end{figure}

In Fig.\ \ref{figS-AHE}, we present the analysis of the field-dependent Hall effect (first column) following the formalism previously applied to $4f$ and $5f$ compounds at low temperatures\ \cite{Yamada1993,NiuPRR2020}. 
Under the assumption that skew scattering (extrinsic) is the dominating contribution to the anomalous Hall effect of \UTe, the Hall resistivity may be proportional to the square of the longitudinal resistivity multiplied by the magnetization in the coherent regime at low temperature\ \cite{Nagaosa2010AHE,Yamada1993}. 
We plot the scaled Hall resistivity $\rho_{xy}/H$ against the product $\rho_{xx}^2 M/H$ scaled by magnetic field (middle panels in Fig.\ \ref{figS-AHE}). 
The data sets, both at low fields ($0\,\mathrm{T}\leq\mu_0H\leq 32\,$T) in the second column and in the high-field range ($45.5\,\mathrm{T}\leq\mu_0H\leq 65\,$T) in the third column, exhibit linear dependencies. 
In order to account for the magnetization at $\theta\approx30^\circ$, we included previously published data from Ran et al. and Miyake et al.\ \cite{Ran2019extreme,Miyake2019}. 
Compared to previous results obtained for $H\parallel b$, the data show a similar linearity indicative for an extrinsic anomalous Hall effect (AHE) component associated with skew scattering
From this analysis, the normal Hall coefficient $R_0$ can be estimated from the intercept, indeed $\rho_{xy}/H=R_0+\rho_{xx}^2M/H$. 
At the lowest temperature (0.6\,K), the resulting value changes from $0.06\,\mu\Omega$cm to about $0.1\,\mu\Omega$cm upon transitioning from below to above \Hm.
This corresponds to a jump by roughly factor of two. 
In comparison, $\rho_{xy}$ changes by a factor of 10 for $H\parallel b$\ \cite{NiuPRR2020}. 
The latter result was associated with a significant Fermi-surface reconstruction at \Hm due to a drastic reduction of the carrier concentration. 
However, other quantities such as  the $A$ coefficient in the temperature-dependent resistivity\ \cite{KnafoJPSJ2019} or the Sommerfeld coefficient (predicted from magnetization results\ \cite{Miyake2019}) do not exhibit similar dramatic jumps. 
Therefore, the result suggests an additional intrinsic AHE component (associated with a topological Berry-curvature contribution), which is independent of the scattering time $\tau$ and, thus, disregarded by the analysis described above. 
Another point, in favor of an enhanced intrinsic anomalous-Hall component, would be that the overall Hall conductivity above \Hm is of the order of $10^4-10^5(\Omega\mathrm{cm})^{-1}$. 
According to what has been empirically determined for various materials (see the review by Nagaosa\ \cite{Nagaosa2010AHE}), this is assumed to be the ``good-metal'' regime, mainly dominated by intrinsic contributions in the Hall effect.

Such intrinsic contributions may be described in terms of the Berry curvature formalism and can arise from various origins, e.g.:
\begin{itemize}
\item Avoided band crossings in the spin-polarized band structure close to or at the Fermi edge\ \cite{Burkov2014},
\item Complex magnetic order involving spin-chirality scattering effects that may cause an additional transverse anomalous velocity\ \cite{Fujishiro2021giant}.
\end{itemize}
We believe that this is an important finding, as we demonstrate here that exactly this enhanced intrinsic Hall component seems to get drastically suppressed at the angular range where the hfSC phase emerges. 
Interestingly, Weyl physics has already been proposed to be of importance in \UTe\ \cite{Shishidou2021TologigicalBand}.
In order to estimate the Berry-curvature effect in momentum space, detailed information on the Fermi surface in the high-field phase are required.
As first magnetic quantum oscillations have already been reported\ \cite{Aoki2022dHvA,Eaton2023}, it may become possible in the near future to obtain details about the effect of \Hm on the band structure of \UTe. 

The general analyses of the Hall effect is complicated by the appearance of metamagnetism in \UTe, the lack of detailed information about the magnetic structure and its effects on the band structure of \UTe.
In order to discriminate the orbital from the anomalous coefficient an approach would be to study the high-field Hall effect in the field range where the magnetism is saturated so that the AHE contribution remains constant. 
As the magnetization of \UTe does not saturate above \Hm up to 70\,T such an approach could not be applied in our case. 

\subsection*{Supplementary Note 5: An attempt to fit \texorpdfstring{$H_\mathrm{c2}$}{Hc2} in the hfSC phase assuming Jaccarino-Peter compensation.}
\label{NoteSCmodel}

Recently, Rosuel et al.\ \cite{RosuelPRX2023} showed that the field-reinforced sc phase close to (below) \Hm for $H\parallel b$ can be described by a field-controlled strong-coupling constant $\lambda(H)$ in the strong-coupling regime.
The model relies on a strong-coupling calculation of \Hc for $s$-wave superconductors, including both orbital and paramagnetic effects. 
The effect of spin-triplet pairing is ``mimicked'' by suppressing the paramagnetic effect. 
The parameters of the model for the determination of the critical temperature are a typical energy scale of the fluctuations responsible for the pairing ($\Omega$), the pairing strength $\lambda$, and a screened Coulomb repulsion ($\mu^{\ast}$). 
The spectral density of interactions is defined in a minimal form\ \cite{Bulaevskii1988Hc2}: 
\renewcommand{\theequation}{S\arabic{equation}}
\begin{equation}
\centering{
    \alpha^2 F(\omega) = \frac{\lambda \Omega}{2}\delta(\omega - \Omega).
}
\end{equation}
\Tc and the \Hc are derived from the solution of the resulting Eliashberg equations.
This also yields a renormalization of the bare averaged Fermi velocity ,$\langle v_\mathrm{F}\rangle$, that controls \Hc in a given direction (see Appendix of Ref.\ \cite{RosuelPRX2023}).

The field dependence of $\lambda$ is required to account for the reinforcement of \Hc for $H \parallel b$.
It is deduced from the comparison between the experimental data and calculations for different values of $\lambda$, which is the most sensitive parameter of the model ($\Omega$ and $\mu^{\ast}$ are chosen to remain field independent).

Two scenarios were proposed in Ref.\ \cite{RosuelPRX2023} for the high-field field-reinforced phase ($H \parallel b$): 
Either a spin-triplet ESP state for the field-reinforced phase, hence, with no paramagnetic limit at all, or a (counter-intuitive) spin-singlet phase, possibly linked to antiferromagnetic fluctuations that develop upon approaching \Hm.
In the second case, the enhancement of the paramagnetic limit, due to the increase of the $\lambda$, plays a major role for the control of \Hc in the field-reinforced phase.
Support for this scenario stems from the strong broadening of the specific-heat anomaly observed in the field-reinforced phase and from the angular dependence of this broadening\ \cite{RosuelPRX2023}.

We extend this model to finite tilt angles by including two simple hypotheses:  
Following Ref.\ \cite{RosuelPRX2023}, $\lambda(H)$ in the field-reinforced phase follows the angle-dependence of \Hm, hence, $\lambda(H/H_\mathrm{m}(\theta))$, with $H_\mathrm{m}\propto 1/\cos\theta)$. 
For $\langle v_\mathrm{F}\rangle$ controlling \Hc, if $\bar{v}_b^0$ and  $\bar{v}_c^0$ are those for fields along the $b$ and $c$ axis, respectively, at an angle $\theta$ between the two we use:
\begin{equation}
\centering{
    \langle v_\mathrm{F}(\theta)\rangle = \sqrt{\left(\bar{v}_b^0 \cos\theta\right)^2 + \left(\bar{v}_c^0 \sin\theta\right)^2}.
}
\label {Eq:vF}
\end{equation}
Here, we assume an isotropic Pauli limit, hence, no angular dependence of $g$ (with an absolute value of $g=0$ in the ESP spin-triplet case or $g=2$ in the spin-singlet case).

We apply this model to the hfSC phase using the same values of $\Omega$, $\mu^{\ast}$, $\bar{v}_b^0$, and $\bar{v}_c^0$ for all scenarios, consistent with Ref.\ \cite{RosuelPRX2023}. 
Therefore, we neglect any effects related to possible changes of the Fermi surface\ \cite{NiuPRL2020} or the characteristic fluctuation energy at \Hm. 
However, we assume that the compensation effect between $H$ and \He, together with the ``distance'' of $H$ from \Hm lead to a new field dependence of $\lambda$.
For the scenario of spin-singlet pairing this will change the paramagnetic limit as well.

This last point requires a determination of the compensation as a function of $H$ and $\theta$. 
At the mean-field level, as mentioned in the main text, \He depends crucially on the $H$ dependence of both the longitudinal and transverse magnetization: 
\He$=J_c<M_c>\hat{c} + J_b<M_b>\hat{b}$, with the anisotropic exchange constants $J_c$ and $J_b$ and the magnetization components $<M_c>$ and $<M_b>$ along the $c$ and $b$ axes, respectively. 
Presently, the dependence of the magnetization on $H$ is unknown: 
Our magnetic-torque measurement and the large jump observed at \Hm indicate that the magnetization jump is not purely longitudinal. 
It is even likely that $<M_b>$ remains close to its value for $H \parallel b$ at \Hm, which is in agreement with the angle dependence of \Hm, following  $1/\cos{\theta}$.
However, the large jump in $M(H)$ at \Hm observed for finite tilt angle in previous works\ \cite{MiyakeA2021, Ran2019extreme} is also indicative of a significant component along $c$, likely to grow further for fields above \Hm. 
Additional experiments are required to determine the details of this dependence. 

\begin{figure}[tb]
	         \centering
         \includegraphics[width=0.65\textwidth]{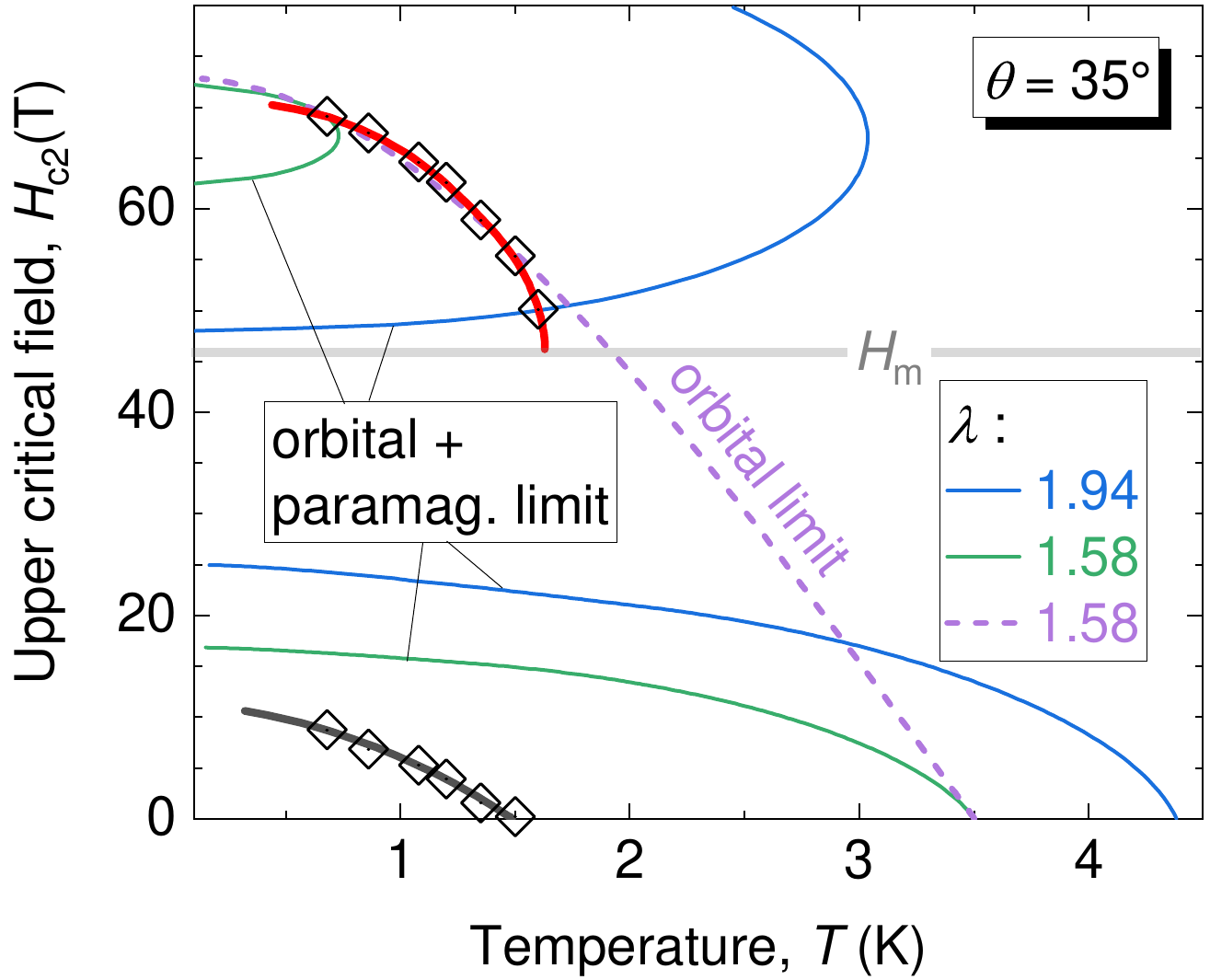}
         \caption{
         \textbf{$H_\mathrm{c2}(T)$ including the Jaccarino-Peter compensation effect\ \cite{Jaccarino1962}.} 
         The black diamonds are \Hc values in the LF and hfSC phases as determined by experiment for $\theta=35^\circ$. 
         Black solid line is an orbital fit using $\lambda(H)$ (presented in the main text in Fig.\ 4b) for the low-field phase, and parameters for the Fermi velocity along $b$ and $c$ axes taken from Ref.~\cite{RosuelPRX2023}. 
         The purple dashed line is a fit in the pure orbital limit with a constant value of $\lambda=1.584$, and the same bare Fermi velocity as in the LF phase. 
         Green and blue solid lines are fits including orbital and paramagnetic limitation with two different $\lambda$ values as indicated, both below \Hm (without \He, i.e., no compensation) and above \Hm, where \He$=-70\,$T (collinear with $H$) partly compensates the external field. 
         The red solid line is the resulting fit using the $\lambda(H)$ (presented in the main text in Fig.\ 4b) for $H\geq$\Hm.
         }
         \label{figS-JPfit}
    \end{figure}

Hence, we have limited the evaluation of this scenario to the \Hc data set at $\theta=35^{\circ}$, where the almost complete suppression of the Hall angle at 70\,T suggests a very good compensation of $H$ by \He. 
Furthermore, we assume that above \Hm, up to 70\,T, \He (linked to the magnetization jump) remains constant, equal to $-70\,$T, and collinear with $H$.
In \UTe, the crystallographic orientation $[011]$ seems specific in many respects: The $(011)$ plane is a natural cleavage plane of single crystals of \UTe; It is associated with a Yamaji magic angle related to the warping of the Fermi Surface; It becomes the new direction for the tetragonal $c$ axis after the recently observed pressure-induced structural transition\ \cite{HondaJPSJ2023}. 
Microscopic models might explain if its importance for the magnetic properties of \UTe is coincidental. 
Perfect compensation at 70\,T means that at this field, \Hc should be completely limited by the orbital effect. 
With $\langle v_\mathrm{F}(\theta)\rangle$ determined as in Eq.~\ref{Eq:vF}, the only parameter left to be adjusted is the value of $\lambda$ at this field. 
Fig.\ \ref{figS-JPfit} shows the data of \UTe at 35$^{\circ}$ for both in the lfSC and hfSC phases, and the orbital \Hc calculated for an $H$-independent value of $\lambda = 1.58$ (dashed purple line).
    
In Fig.\ \ref{figS-JPfit}, we also show the result of the same calculation with $\lambda = 1.58$ including the paramagnetic limit (solid green line), controlled by $g \mu_\mathrm{B} (H - H_\mathrm{ex})$, where \He is zero below \Hm and -70\,T above \Hm.  
The reentrant hfSC phase then can emerge above \Hm thanks to the compensation by \He. 
This is clearly visible for both $\lambda$ values shown. 
For the \Hc value closest to \Hm we found $\lambda=1.94$ (blue solid line in Fig.\ \ref{figS-JPfit}). 
Below \Hm, we find that the maximum of \Hc for both $\lambda$ values, namely $H_\mathrm{c2}(0) \approx 17\,$T and $H_\mathrm{c2}(0) \approx 25\,$T, respectively, lie below the $H$ values necessary to reach the corresponding $\lambda(H)$ (see Fig.\ 4b in the main text: $H(\lambda=1.58) \approx 22\,$T or $H(\lambda=1.94)\approx 32\,$T. 
As a consequence, SC is suppressed by the paramagnetic limit. 
The reason is that \Hm increases approximately with $1/\cos(\theta)$ and below \Hm the $H$-induced increase of $\lambda (H/H_\mathrm{m})$ at $35^\circ$ is too small to prevent the paramagnetic limit from suppressing the superconducting state. 
The complete $\lambda(H)$ curve above \Hm at $35^\circ$ in Fig.\ 4b is deduced from a smooth fit of the values of $\lambda$ required to reproduce the data within both scenarios (Jaccarino-Peter compensation shown in Fig.\ \ref{figS-JPfit} and pure orbital limitation shown in Fig.\ 2e).
The overall $H$ dependence of $\lambda$ resembles that of the specific heat ($C_p/T$) in the normal state around \Hm for field along the $b$ axis\ \cite{RosuelPRX2023}. 
It differs, however, from the positive jump that was proposed from analysis of the magnetization by the Clausius-Clapeyron relations\ \cite{MiyakeA2021}. 
As explained in Ref.\ \cite{RosuelPRX2023}, $C_p/T$ in \UTe even at temperatures as low as 1.8\,K is rather complex and consists more contributions than that of the Sommerfeld coefficient. 
Thus, it is difficult to draw a direct connection between the $H$ dependence of $\lambda$ and that of $C_p/T$ (through the mass renormalization induced by pairing interactions). 
Moreover, at \Hm, many phenomena can come into play, e.g, a change of the band structure (Fermi-surface topology), which are not directly related to $\lambda$. 
\end{document}